\documentclass[10pt,conference,letterpaper]{IEEEtran}
\pdfoutput=1
\usepackage{balance}
\usepackage{url}
\usepackage{epsfig, amssymb, amsmath, bbm, multirow, graphicx}
\usepackage[caption=false,font=footnotesize]{subfig}
\usepackage[linesnumbered,ruled]{algorithm2e}

\SetCommentSty{mycommfont}

\newcommand{\nop}[1]{}

\newtheorem{theorem}{\bf Theorem}

\newtheorem{example}{\bf Example}

\DeclareMathOperator*{\argmax}{arg\,max}

\title{On Efficiently Detecting Overlapping Communities over Distributed Dynamic Graphs}

\author{%
{Xun Jian{\small $~^{\#}$}, Xiang Lian{\small $~^{*}$}, Lei Chen{\small $~^{\#}$} }%
\vspace{1.6mm}\\
\fontsize{10}{10}\selectfont\itshape
$^{\#}$\,Hong Kong University of Science and Technology, Hong Kong, China\\
\fontsize{9}{9}\selectfont\ttfamily\upshape
\{xjian, leichen\}@cse.ust.hk%
\vspace{1.2mm}\\
\fontsize{10}{10}\selectfont\rmfamily\itshape
$^{*}$\,Kent State University, Ohio, USA\\
\fontsize{9}{9}\selectfont\ttfamily\upshape
xlian@kent.edu
}

\begin{document}

\maketitle

\begin{abstract}
Modern networks are of huge sizes as well as high dynamics, which challenges the efficiency of community detection algorithms. In this paper, we study the problem of overlapping community detection on distributed and dynamic graphs. Given a distributed, undirected and unweighted graph, the goal is to detect overlapping communities incrementally as the graph is dynamically changing. We propose an efficient algorithm, called \textit{randomized Speaker-Listener Label Propagation Algorithm} (rSLPA), based on the \textit{Speaker-Listener Label Propagation Algorithm} (SLPA) by relaxing the probability distribution of label propagation. Besides detecting high-quality communities, rSLPA can incrementally update the detected communities after a batch of edge insertion and deletion operations. To the best of our knowledge, rSLPA is the first algorithm that can incrementally capture the same communities as those obtained by applying the detection algorithm from the scratch on the updated graph. Extensive experiments are conducted on both synthetic and real-world datasets, and the results show that our algorithm can achieve high accuracy and efficiency at the same time.
\end{abstract}

\vspace{-1ex}
\section{Introduction}
In many real-world applications, we often model the underlying data as graphs. For example, we can consider the World Wide Web as a graph by treating each single page as a vertex and hyper-links between pages as edges. Similarly, a social network can also be represented by a graph where users are vertices and friend relationships between users are edges. It has been shown that many real-world networks have a significant property of community structure \cite{girvan2002community}, which means that the vertices in the network can be partitioned into communities, such that vertices within a community are densely connected and vertices from different communities are sparsely connected. For example, communities exist in World Wide Web \cite{albert1999internet}, where pages belonging to the same community (topic) are closely connected. Similarly, in social networks \cite{scott2017social}, people sharing similar relationships (e.g., friends) are closely connected. Moreover, communities (especially those in social networks) are usually overlapping with each other \cite{palla2005uncovering}, i.e., different communities can share some vertices, and one vertex can belong to several communities simultaneously. For example, in real social networks, a person can belong to several communities like friends, families, colleagues, neighbors, and so on.



Although the existence of community structure is widely recognized, there is no universally accepted definition of what a community is. Various algorithms have been proposed in recent years, trying to detect communities from different perspectives. Some prior works \cite{palla2005uncovering,kumpula2008sequential} treat a community as a set of cliques (complete subgraphs). Zhang et al. \cite{zhang2007identification} map vertices into a Euclidean space and then use fuzzy C-Means to cluster the vertices. Ren et al. \cite{ren2009simple} use a probabilistic model and Expectation-Maximization (EM) method to calculate the probability that each vertex belongs to each community. Lancichinetti et al. \cite{lancichinetti2009detecting} optimize a fitness function locally to add/remove vertices into/from a community gradually. Zhang et al. \cite{zhang2009parallel} compute a topology graph and the propinquity (or similarity) between vertices iteratively, then extract communities from a relatively stable topology graph. Xie et al. \cite{xie2012towards} propose an algorithm to propagate labels between vertices and let communities emerge as popular labels.

Most of the previous works mentioned above share a common property that is they all aim at detecting communities on a static graph or on a single machine. Some works like \cite{zhang2009parallel} propose solutions for distributed static graphs. Some other works like \cite{cazabet2010detection,xie2013labelrankt} detect communities on dynamic graphs on a single machine. However, many real-life networks are both of large scale and dynamic. For example, most popular social networks (e.g., Facebook, Whatsapp and Wechat) have reached hundreds of millions or even billions of users in June, 2017 \cite{techcrunch}. Moreover, there are over 4 billion pages on the Internet in June, 2017 \cite{wwwsize}. Identifying communities in such big networks requires huge storage and computation resources which are expensive to be obtained from a single machine. Instead, distributed storage and computing clusters can provide sufficient resources in a relatively easy and cheap way.

Other than the large scale, social networks and the Internet change rapidly. Vertices and edges are added and deleted from time to time. Note that, it is not efficient to treat the updated graph as a new graph, and re-run the community detection algorithm from scratch. Thus, it is important to incrementally monitor the evolution of the communities, upon graph updates.


Existing community detection algorithms over dynamic graphs have the following shortcomings. The iLCD algorithm \cite{cazabet2010detection} cannot handle edge/vertex deletions and LabelRankT \cite{xie2013labelrankt}, cannot guarantee the result given by incremental updating is of equal quality compared to the result calculated from scratch. To the best of our knowledge, no prior works can detect overlapping communities over distributed and dynamic graphs accurately.

In this paper, we focus on incremental community detection over distributed and dynamic binary graphs (a binary graph refers to a graph with no weights or directions on its edges). Any network can be transformed to a binary graph by removing the directions of edges and applying thresholding on weighted edges. We first look into an efficient and highly parallelizable algorithm called SLPA \cite{xie2012towards} for static graphs. By smoothing the voting result with uniform-picking, we propose the \textit{randomized Speaker-Listener Algorithm} (rSLPA) that is able to incrementally detect communities with high efficiency. Then we discuss the time complexity of the proposed incremental algorithm, as well as the best and the worst cases. Experimental results show that rSLPA can handle dynamic graphs efficiently and effectively.

Specifically, we make the following contributions.
\begin{itemize}
\item    We design an efficient algorithm rSLPA that can incrementally detect communities over distributed and dynamic graphs.
\item    We derive the expected complexity of incremental updating of rSLPA, as well as the upper and lower bounds.
\item    We verify the effectiveness and efficiency of rSLPA by conducting extensive algorithms over both synthetic and real-world datasets.
\end{itemize}

The rest of this paper is organized as follows. In Section \ref{sec:prelim} we introduce the community detection problem, as well as the details of SLPA. Then in Section \ref{sec:smoothing} we describe our approach in both label propagation stage and post-processing stage. In Section \ref{section_incremental}, we analyze the possible situations brought by graph changes, and then propose an algorithm to incrementally update the result. Analysis to the complexity and extreme cases are given as well. Extensive experiments are conducted in Section \ref{section_exp} to show the effectiveness and efficiency of our algorithm. Finally, we discuss the related works in Section \ref{sec:related} and conclude this paper in Section \ref{sec:conclusion}.

\vspace{-2ex}
\section{Preliminaries}
\label{sec:prelim}

\subsection{The Community Detection Problem}
Given a binary graph $G(V, E)$, the community detection problem is to output a set of communities, each containing a set of vertices. Vertices within one community should be densely connected, and that from different communities should be sparsely connected.

Since there is no globally accepted definition of a community, most of the existing algorithms are either based on heuristics or maximizing an objective \cite{ren2009simple,lancichinetti2009detecting,zhang2009parallel}. In this paper, we consider one of the heuristics, label propagation. In this approach, labels are propagated among vertices, and vertices that receive similar information after enough iterations are assigned to the same community. We choose this heuristic due to the following two facts: (1) some prior works \cite{xie2012towards,kuzmin2013parallel} show that this approach can achieve high accuracy as well as high efficiency, and (2) the most widely used objective Modularity has some limitations \cite{lancichinetti2011limits}. In the following, we further describe the label propagation process in the SLPA algorithm.

\subsection{The SLPA Algorithm}
The \textit{Speaker-Listener Label Propagation Algorithm} (SLPA) \cite{xie2012towards} is a natural algorithm simulating label propagation among vertices. Compared to other community detection algorithms, SLPA has a relatively low complexity and competitive performance on real datasets \cite{harenberg2014community}. Benefiting from the label propagation model, SLPA can be easily parallelized \cite{kuzmin2013parallel} and applied on distributed graphs.

Given a graph $G(V, E)$, each vertex $v_i \in V$ is associated with a sequence of labels $L_i=(l_i^0, l_i^1, \ldots, l_i^k)$. At the beginning each $L_i$ is empty, and the labels are assigned as the algorithm runs. SLPA can be summarized by three main stages below:
\begin{enumerate}
\item Initialization;
\item Label Propagation; and
\item Thresholding.
\end{enumerate}

In the \textit{Initialization} stage, each vertex is assigned with a unique label (usually the ID of this vertex), which means $L_i=(l_i^0)=(i)$. Later, the label sequence of each vertex will be enlarged in the label propagation stage.

\begin{figure}[htbp]
\centering
\includegraphics[width=0.9\linewidth,keepaspectratio]{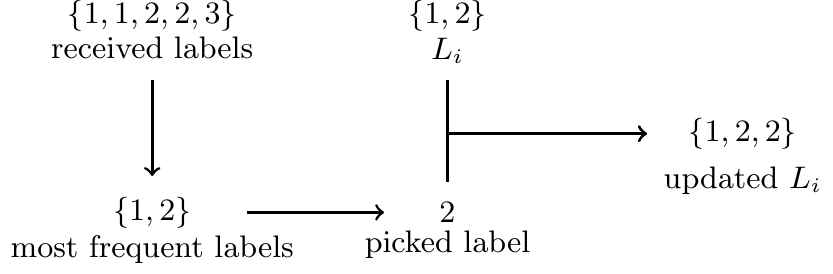}\vspace{-2ex}
\caption{A label selection example. Two labels have the same
(largest) frequency, so one of them is uniformly
picked.}\vspace{-2ex} \label{fig:label_selection}
\end{figure}

The \textit{Label Propagation} stage contains two super-steps: \textit{label sending} and \textit{label selection}. In the \textit{label sending} step, each vertex sends a label to each of its neighbors. Typically, for each vertex $v_i$ and each of its neighbor $v_j$, $v_i$ will uniformly pick a label from $L_i$ for sending to $v_j$. After this step, every vertex will receive a set of labels from its neighbors. Let $N_i$ be the neighbor set of $v_i$, and $M_i$ be the set of labels received by $v_i$, where $|N_i|=|M_i|$. In the \textit{label selection} step, each vertex $v_i$ will pick the most frequent label in $M_i$, and add it to $L_i$. If there are several most frequent labels (i.e., with the same frequency), a random label will be picked uniformly. Figure \ref{fig:label_selection} shows a typical example where both labels $1$ and $2$ are the most frequent. In this example label $2$ is finally picked and appended to the label sequence.

The two super-steps will be performed iteratively for $T$ times (In the original paper \cite{xie2012towards}, $T$ is set to 100.), to let each label sufficiently be propagated on the graph. After $T$ times, each vertex has a label sequence of size ($T+1$).

Finally, in the \textit{Thresholding} stage, labels in each $L_i$ that are below a frequency threshold $\tau$ will be filtered out, and the remaining labels in each label sequence indicate the communities that this vertex belongs to.

\vspace{-1ex}
\section{Smoothing voting in SLPA}
\label{sec:smoothing}
In this section, we first discuss the details of the voting process in SLPA. Taking the graph dynamics into consideration, where the neighbors of each vertex may change, we analyze the diverse results of this voting process when the neighbors change a bit. Then we introduce a pure random process to replace the voting process in SLPA, which can be treated as a smoothing to the original one. We show that by applying this new process, the complexity of the algorithm is reduced, and more importantly, the results become trackable so that we are able to design an incremental updating algorithm over dynamic graphs in Section \ref{section_incremental}. Then we propose a post-processing approach to extract communities from the label propagation results. Later in Section \ref{section_exp}, experiments show that our new strategy can give efficient and competitive results compared to the original algorithm over static and dynamic graphs.

\vspace{-1ex}
\subsection{Smoothing the Voting Process}
\label{subsec:uniform}
The basic idea behind the voting process of SLPA is to make densely-connected vertices ``agree'' on a common label within a limited number of iterations. Intuitively, within a community, popular labels will appear more frequently than other labels in the voting process, and thus have higher chances to win the voting and become more popular. This property, which we call \textit{concentration}, allows communities to arise as popular labels. Meanwhile, popular labels in one community are not likely to be propagated to other communities. This is because the vertices on the ``boundary'' of two communities are sparsely connected, which makes popular labels on one side become unpopular on the other side. This property, which we call \textit{trapping}, will prevent communities from mixing together.

Although the voting process achieves good effectiveness in the experiments of the original work, it also makes the results very sensitive to the popularity of each label and almost unpredictable, as shown in Example \ref{exa:voting_result}. When the graph structure changes, this issue makes it very hard to incrementally update the voting results, instead of running the algorithm from scratch.

\begin{figure}[htbp]
    \centering\vspace{-2ex}
    \subfloat[3 voters: $(1,2)$, $(1,2)$, and $(1,1)$]{
        \includegraphics[width=.45\linewidth, keepaspectratio]{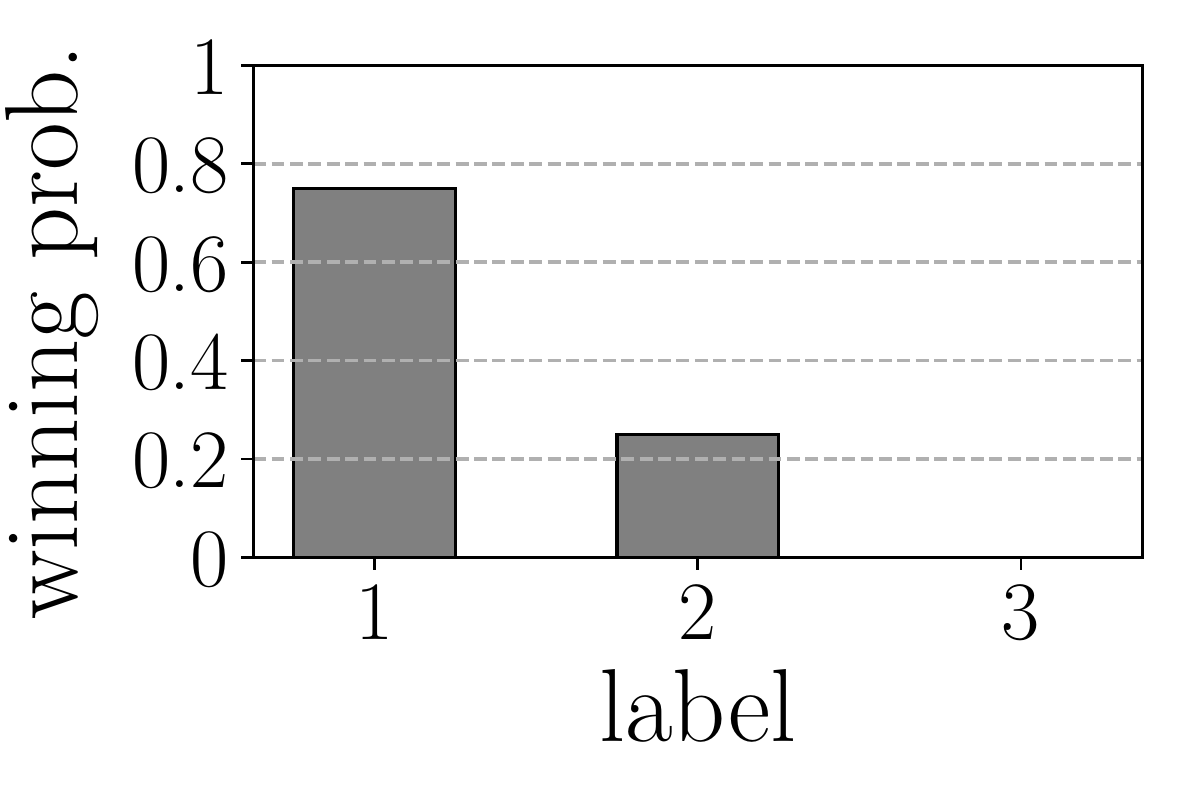}
        \label{fig:voting_diverse_a}
    }%
    \hspace{0.02\linewidth}
    \subfloat[3 voters: $(1,2)$, $(1,2)$, and $(1,3)$]{
        \includegraphics[width=.45\linewidth, keepaspectratio]{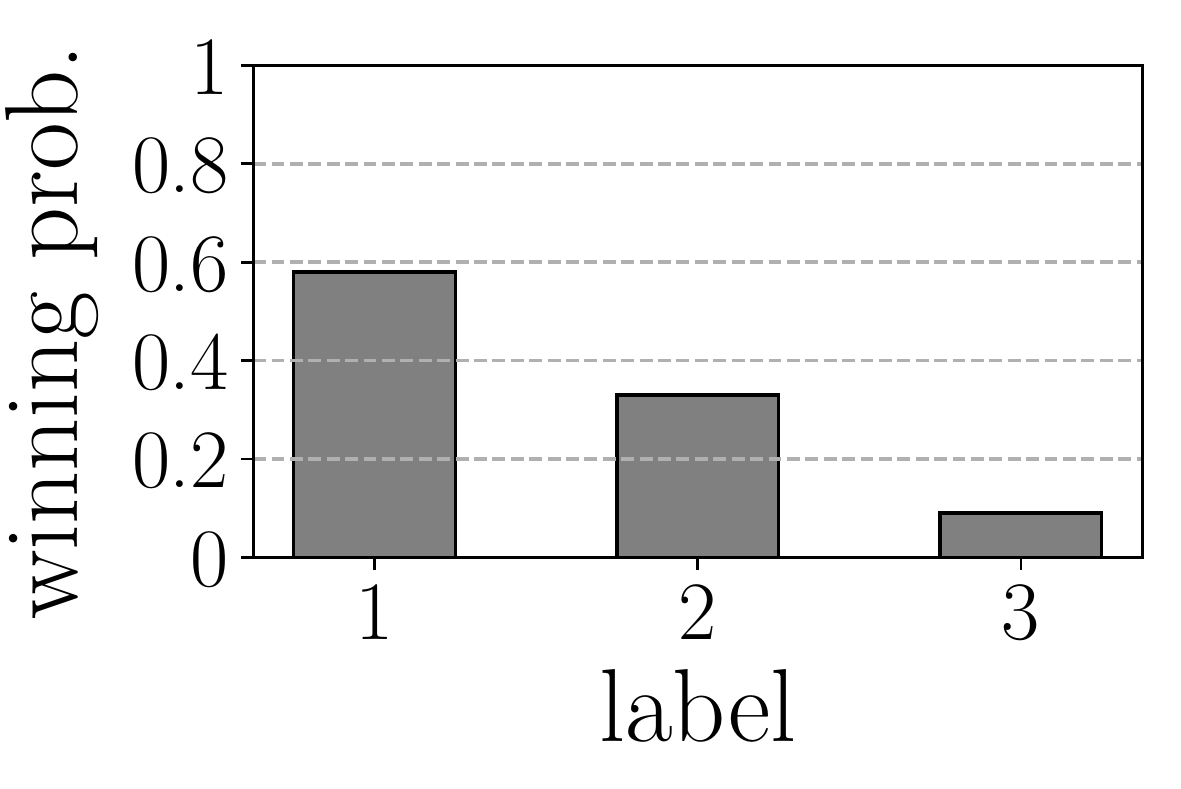}
        \label{fig:voting_diverse_b}
    }%
    \hspace{0.02\linewidth}
    \subfloat[3 voters: $(2,2)$, $(1,1)$, and $(1,1)$]{
        \includegraphics[width=.45\linewidth, keepaspectratio]{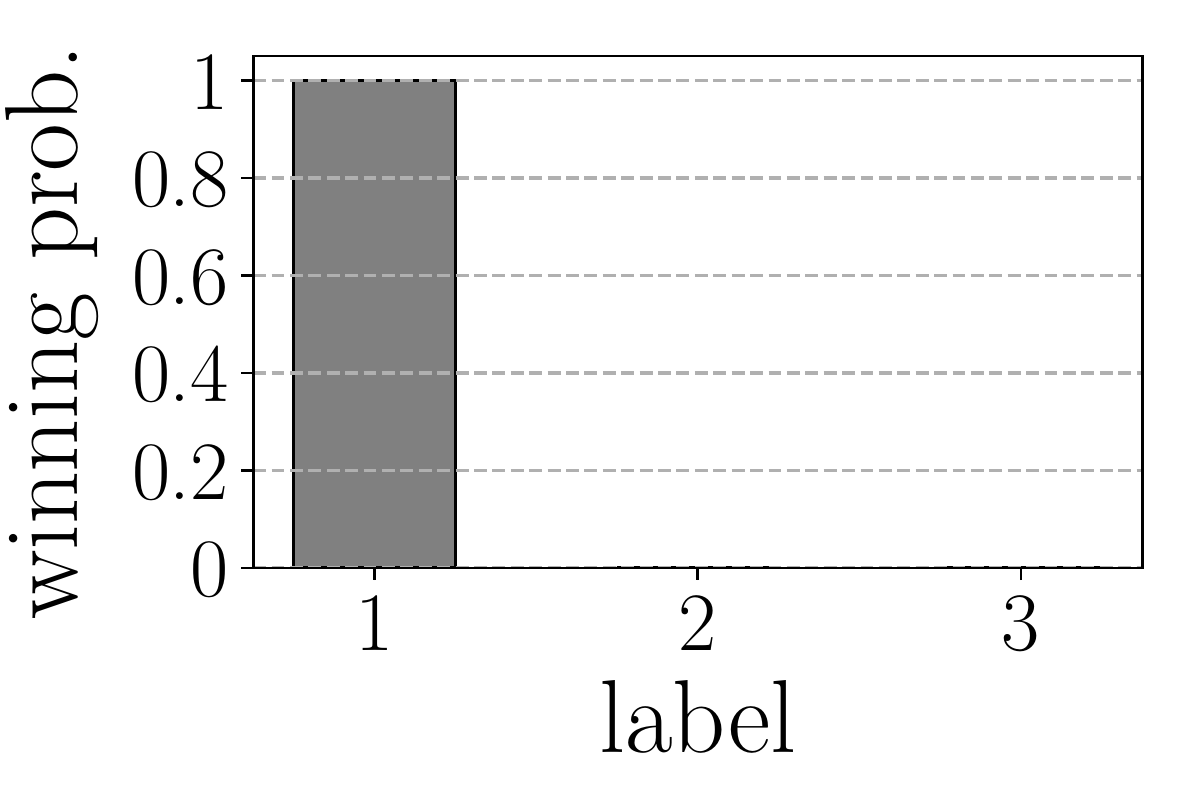}
        \label{fig:voting_diverse_c}
    }%
    \hspace{0.02\linewidth}
    \subfloat[2 voters: $(2,2)$, $(1,1)$]{
        \includegraphics[width=.45\linewidth, keepaspectratio]{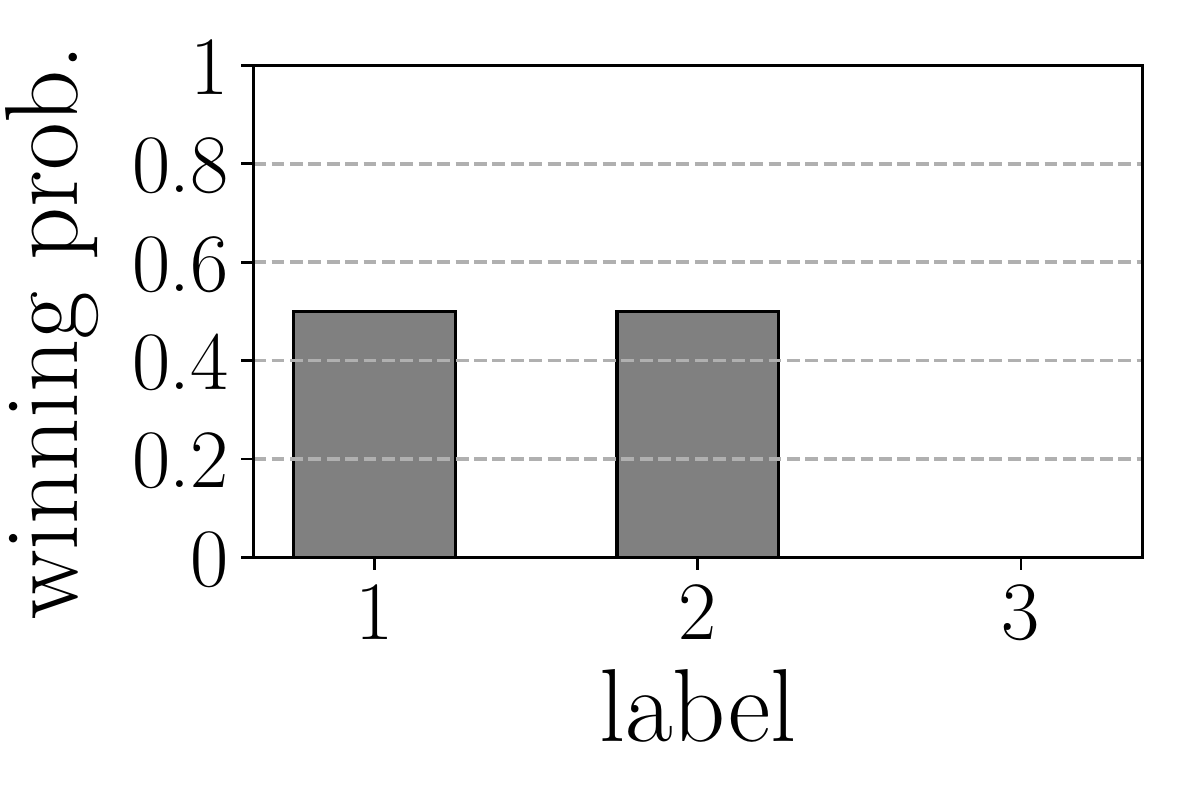}
        \label{fig:voting_diverse_d}
    }\vspace{-1ex}
    \caption{Probability distribution of labels to win the plurality voting, each with different voters.}\vspace{-2ex}
    \label{fig:voting_diverse}
\end{figure}

\begin{example}
\label{exa:voting_result}
In Figure \ref{fig:voting_diverse} there are four results of plurality voting under different settings. In Figure \ref{fig:voting_diverse_a}, three voters are with label sequences $(1,2)$, $(1,2)$ and $(1,1)$, respectively. Under this setting, label $1$ will win the vote with the highest probability, and label $2$ has a much lower probability to win. Other figures can be interpreted in the same way.

Compared to Figure \ref{fig:voting_diverse_a}, in Figure \ref{fig:voting_diverse_b} we only change the label sequence of the third voter from $(1,1)$ to $(1,3)$. Intuitively, the winning probability of label $1$ should decrease and that of label $3$ should increase, because of the population change. The figure does confirm our prediction, but surprisingly the winning probability of labels $2$ drops though we did nothing to it. This case implies that any changes to one single label may have influence on any other labels.

Also compared to Figure \ref{fig:voting_diverse_a}, in Figure \ref{fig:voting_diverse_c} we exchange two labels between the first two voters, so that label $2$ only exists in one sequence. Although the population of each label is not changed, the winning probability distribution changes dramatically, which shows the discreteness of the voting process.

In Figure \ref{fig:voting_diverse_d} we remove the third voter in Figure \ref{fig:voting_diverse_c}. As we can see, the winning probability of label $2$ suddenly increases from 0 to 0.5. This extreme case shows that the result can be very sensitive to the change of the absolute number of voters.
\end{example}

To overcome this issue, we consider another uniform-picking process, which is, given $M_i$, label $l$ will be uniformly picked from $M_i$ as the result. Using this process, the probability that $l$ equals $l_k$ will be proportional to the frequency of $l_k$ in $M_i$. Thus, popular labels will have higher probability to be picked, and the \textit{concentration} property can be achieved. On the other hand, labels will have a small chance to be propagated through a sparse connected area for the same reason, thus the property of \textit{trapping} can also be achieved.

\begin{figure}[htbp]
    \centering\vspace{-2ex}
    \subfloat[Original voting]{
        \includegraphics[width=.45\linewidth, keepaspectratio]{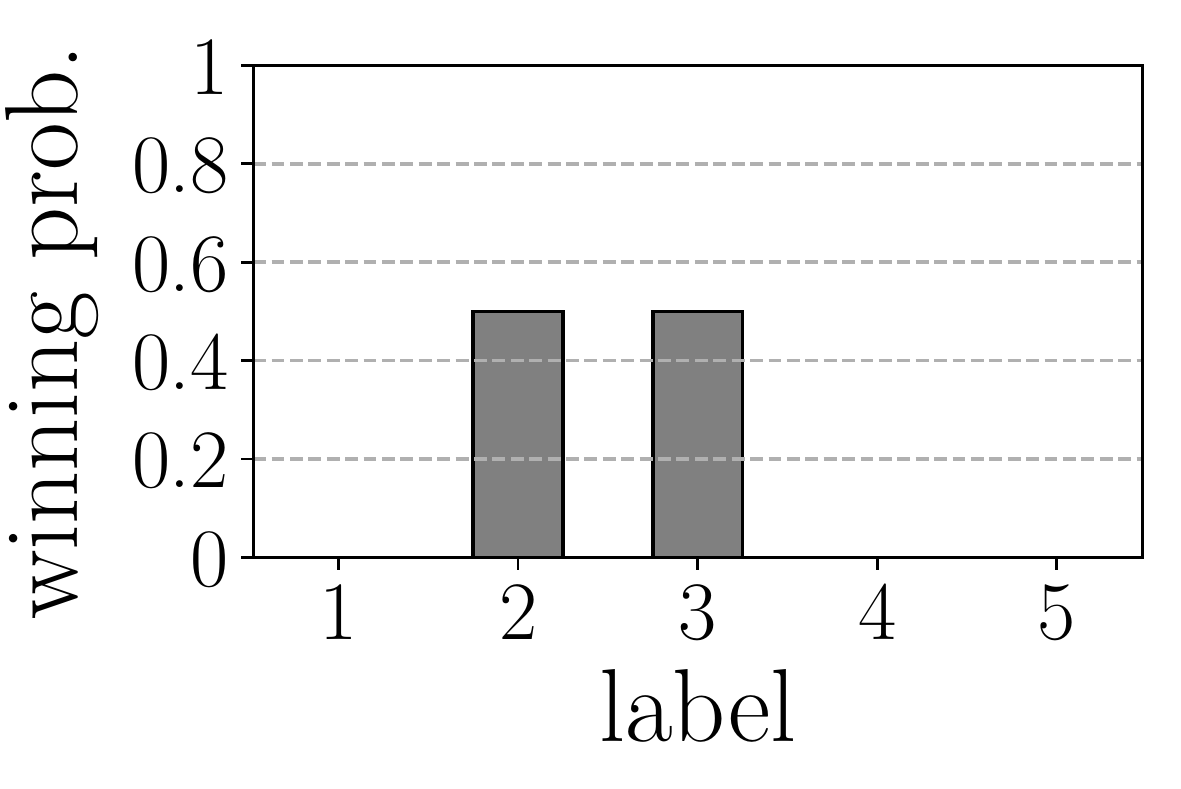}
    }%
    \hspace{0.02\linewidth}
    \subfloat[Uniform-picking]{
        \includegraphics[width=.45\linewidth, keepaspectratio]{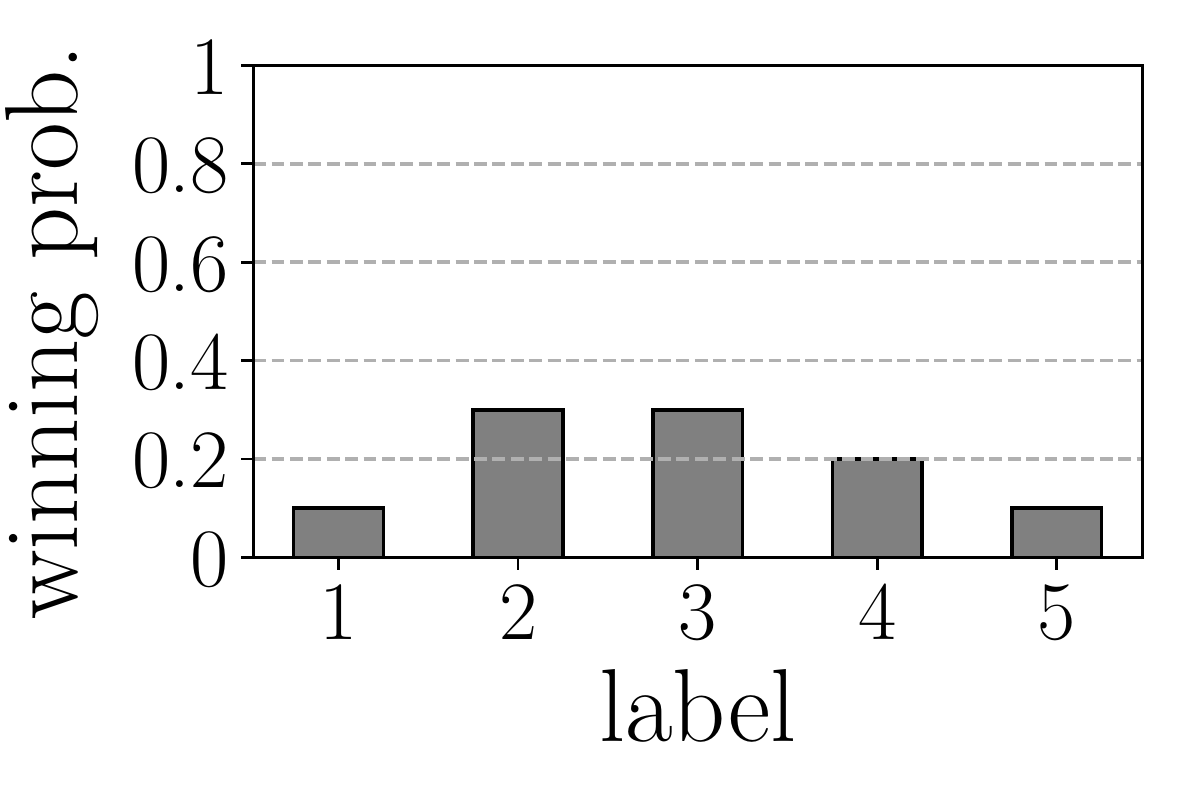}
    }
    \caption{Probability distribution of labels to win, where $M_i=(1,2,2,2,3,3,3,4,4,5)$.}\vspace{-2ex}
    \label{fig:compare_voting_random}
\end{figure}

The major difference between this new process and the old one is that, using the new process, the probability distribution of the result is more ``smooth''. With the old voting process, the most popular label(s) will share the same probability to be picked, and other labels all have probability $0$ to be picked, in other words, the probability distribution only has two levels. However, with the new process, each label will be picked with probability proportional to its population in $M_i$, which means there can be more than two levels. In Figure \ref{fig:compare_voting_random} we show a typical example comparing the result distribution of these two processes, where we can see the probability distribution of uniform-picking has a smoother shape.

Another thing that should be noted in Figure \ref{fig:compare_voting_random} is that, the probability distribution of uniform-picking is more ``flat'' than the voting one. Actually given any $M_i$, the highest probability of picking a label by voting is always greater than or equal to the highest probability of picking a label by uniform-picking, which we prove in Theorem \ref{theorem:flat}. Because of this, the label set of each vertex we get is so different from the one we get in SLPA that the original post-processing cannot be applied to select communities. Another procedure is carefully designed to extract communities and we will discuss the details in Section \ref{subsec:post}.

\begin{theorem}
    \label{theorem:flat}
    Let $\mathbb{P}_v(l)$ be the probability that label $l$ is picked by voting, and $\mathbb{P}_u(l)$ be the probability that label $l$ is picked by uniform-picking. Given any $M_i$, the following inequality holds:
    \begin{align*}\vspace{-2ex}
        \max \mathbb{P}_u(l) \leq \max \mathbb{P}_v(l).\vspace{-2ex}
    \end{align*}
    \begin{proof}
        Suppose $|M_i|=n$ and $k$ labels in $M_i$ have the largest population $m$. Apparently, we have:
        \begin{align*}\vspace{-2ex}
            \max \mathbb{P}_u(l)&=\frac{m}{n} \text{ and } \max \mathbb{P}_v(l)=\frac{1}{k}.\vspace{-2ex}
        \end{align*}
        Since the total number of those $k$ labels cannot be larger than $n$, we have:
        \begin{align*}\vspace{-2ex}
            k \cdot m \leq n,\vspace{-2ex}
        \end{align*}
        and
        \begin{align*}\vspace{-2ex}
            \mathbb{P}_u(l_0) = \frac{m}{n} \leq \frac{1}{k} = \mathbb{P}_v(l_0).\vspace{-2ex}
        \end{align*}
    \end{proof}
\end{theorem}

Taking one step further, since $M_i$ is generated by uniformly picking one label from each neighbor's label sequence, the selected label can also be treated as uniformly picked from $U_i=\cup L_j, j \in N_i$, which is the union of all neighbors' label sequences, and the probability of picking $l_k$ will be proportional to the frequency of $l_k$ in $U_i$. We show the proof in Theorem \ref{theorem:union_prob}. Utilizing this property, we can change the \textit{Label Propagation} stage to the following equivalent process:
\begin{enumerate}
    \item[ ] In iteration $k$:
    \item Each vertex $v_i$ picks a neighbor $src_i \in N_i$ and a position $pos_i \leq k$, both uniformly;
    \item Each vertex $v_i$ add the $pos_i$th label in $L_{src_i}$ in $L_i$.
\end{enumerate}
Each label picked through this process also satisfies the probability distribution we talked before, as shown in Theorem \ref{theorem:random_prob}, but the complexity is noticeably reduced. Since each vertex only fetches one label from one neighbor instead of requiring the whole $M_i$, the total number of labels passing through the graph is reduced from two per edge to one per vertex. Thus, the communication cost in each iteration is reduced from $O(|E|)$ to $O(|V|)$. We then summarize the new label propagation algorithm in Algorithm \ref{alg:rslpa}, where the fetching of a label consists of sending request $pos$ and $receiver$ to $src$, and waiting $src$ to send back the label.

\begin{theorem}
    \label{theorem:union_prob}
    Suppose $\{L_1, L_2, \ldots, L_n\}$ is a set of $n$ label sequences, where each sequence has size $m$, and $M$ is generated by uniformly picking one label from each of those sequences. The number of a specific label $l$ in $L_i$ is denoted by $f(l,i)$. If we uniformly pick $r$ from $M$, then for any label $l$ we have:
    \begin{align*}\vspace{-2ex}
        \mathbb{P}(r=l)=\frac{\sum_{i=1}^n f(l,i)}{nm},\vspace{-2ex}
    \end{align*}
    where $\sum_{i=1}^n f(l,i)$ is exactly the frequency of $l$ in $\cup L_i$.
    \begin{proof}
        We use $M(k)$ to denote the label in $M$ which is picked from $L_k$. Since $M(k)$ is uniformly picked from $L_k$, it holds that:
        \begin{align*}\vspace{-2ex}
            \mathbb{P}(M(k)=l)=\frac{f(l, k)}{m}.\vspace{-2ex}
        \end{align*}
        Then we have:
        \begin{align*}\vspace{-2ex}
            \mathbb{P}(r=l)&=\sum_{k=1}^n\mathbb{P}(r=M(k))\cdot\mathbb{P}(M(k)=l) \\
                           &=\sum_{k=1}^n\frac{1}{n}\cdot\frac{f(l,k)}{m} \\
                           &=\frac{\sum_{i=1}^n f(l,i)}{nm}.\vspace{-2ex}
        \end{align*}
    \end{proof}
\end{theorem}

\begin{theorem}
    \label{theorem:random_prob}
    Suppose $\{L_1, L_2, \ldots, L_n\}$ is a set of $n$ label sequences, where each sequence has size $m$. We use $f(l,i)$ to denote the population of label $l$ in $L_i$, and $l_i^k$ to denote the $k$th label in $L_i$. If we uniformly pick $src$ from $\{1, 2, \ldots, n\}$, and $pos$ from $\{1, 2, \ldots, m\}$, then for any label $l$, we have:
    \begin{align*}\vspace{-2ex}
        \mathbb{P}(l_{src}^{pos}=l)=\frac{\sum_{i=1}^n f(l,i)}{nm}.\vspace{-2ex}
    \end{align*}
    \begin{proof}
        Picking the $src$ and $pos$ of a label are two independent random process, thus the probabilities can be simply multiplied together. Then the probability can be obtained by averaging over all possible $src$:
        \begin{align*}\vspace{-2ex}
            \mathbb{P}(l_{src}^{pos}=l)&=\sum_{i=1}^n\mathbb{P}(src=i)\cdot\mathbb{P}(L_i^{pos}=l) \\
                                       &=\sum_{i=1}^n\frac{1}{n}\cdot\frac{f(l,i)}{m} \\
                                       &=\frac{\sum_{i=1}^n f(l,i)}{nm}.\vspace{-2ex}
        \end{align*}
    \end{proof}
\end{theorem}

\begin{algorithm}
    \caption{Randomized Label Propagation}
    \label{alg:rslpa}
    \SetKwInOut{Input}{Input}\SetKwInOut{Output}{Output}
    \SetKwFunction{RandomPick}{RandomPick}
    \SetKwFunction{Fetch}{FetchLabel}
    \SetKwFunction{Append}{Append}
    \SetKwFunction{Thresholding}{Thresholding}
    \SetKwProg{Mapper}{Mapper}{:}{}
    \SetKwProg{Reducer}{Reducer}{:}{}
    \Input{$G(V, E)$, $T$.}
    \Output{A set $L$ where each $L_i \in L$ is the label sequence of $v_i$.}
    \BlankLine
        Initialize $L_i \leftarrow (i)$\;
    \For{$\; t \leftarrow 1$ \KwTo $T \;$}{
        \Mapper{for $v_i$}{
            $src \leftarrow \RandomPick{$N_i$}$\;
            $pos \leftarrow \RandomPick{$\{0,1,\ldots,t-1\}$}$\;
            Emit $(src, pos, i)$\;
        }
        \Mapper{for $v_i$}{
            \ForEach{received $\; (pos, j) \;$}{
                Emit $(j, l_i^{pos})$\;
            }
        }
        \Reducer{for $v_i$}{
            \Append{$L_i$, received $l$}\;
        }
    }
\end{algorithm}

\vspace{-2ex}
\subsection{Post-processing}
\label{subsec:post}
As we discuss in Section \ref{subsec:uniform}, for any given $M_i$, the uniform-picking process will always give a more ``flat'' result than the original one. This will bring two issues on the final result (i.e., $L_i$ of each vertex). First, instead of agreeing on a single frequent label, one community may agree on a similar distribution of labels. Since the uniform-picking process allows each label to be picked with some probability, one community is not likely to have one common frequent label, if two or more labels have close probability to be picked. Actually, several labels are going to be kept as a similar distribution among one community. For example, within a community, some vertices may have labels $(1,1,2,2,3,4,5)$, and other vertices may have labels $(1,1,2,2,3,3,3)$. In such a situation, if we set $\tau=2$, we will get two identical communities (for labels $1$ and $2$) and one broken community (for label $3$). If we set $\tau=3$, we will only get part of this community. Second, in a macro community, the common distribution may change gradually from one sub-area to another. For example, the frequent labels may change from $\{1, 2, 3\}$ to $\{2, 3, 4\}$ and finally to $\{3, 4, 5\}$. Since the difference between two close vertices is small, it is better to assign all the vertices to a single community.

Because of the the first effect that labels within the same community agree on a similar distribution, we cannot directly filter popular labels and extract communities. The similarity between two distributions needs to be carefully calculated and used to identify vertices that belong to the same community. Meanwhile, the gradual changing of common distributions should be captured to find macro communities. Combining those two thoughts, we propose a new post-processing algorithm to extract communities from the label sequences produced by Algorithm \ref{alg:rslpa}.

First, for each edge $e_{ij}$ in the graph, we assign a weight $w_{ij}$ to be the similarity between $L_i$ and $L_j$. Considering that each label sequence is a sample of labels the vertex can get from its neighbors, we use $\mathbb{P}(l_i=l_j)$, the probability of getting the same label from $L_i$ and $L_j$, as the similarity metric. Another advantage of this metric is its simplicity of calculating as it can be obtained by just counting the common labels of two sequences.

After this, we use two thresholds to extract communities. As we discuss above, adjacent vertices that strongly belong to the same community should be very similar to each other, so we use the first threshold $\tau_1$, to filter out edges of low weight, then each connected component (with at least two vertices) on the filtered graph is considered to be a community. In addition, there will be some isolated vertices, which have slightly lower similarity to its neighbors, which means they only belong to some communities weakly. In order to identify those belongingness, we apply the second smaller threshold $\tau_2$ between the isolated vertices and their neighbors. If an isolated vertex $v_i$ is connected to a non-isolated one $v_j$, and the edge weight $w_{ij} \geq \tau_2$, and then $v_i$ is considered to belong to the community that contains $v_j$. Two communities will overlap when some vertices belong to both of them weakly.

The values of $\tau_1$ and $\tau_2$ will heavily influence the community numbers and sizes, as larger thresholds will split the graph into smaller and more sub-graphs. In this paper we use two principles to help decide these two parameters respectively.

The first principle is, maximizing the information. We hope that the extracted communities reveal the connectivity of different areas of the graph, thus, we do not want to get too many micro communities, or only a few macro communities. This is because both situations provide very limited information about the graph. When we use only $\tau_1$ to extract strongly connected communities, large $\tau_1$ leads to many micro communities, and small $\tau_1$ results in a few macro communities. We use the information entropy w.r.t. the relative size of extracted communities to measure the information we get, which is defined as
\begin{align}\vspace{-2ex}
    entropy = -\sum_i \frac{|C_i|}{|V|} \cdot \log \frac{|C_i|}{|V|}, \nonumber\vspace{-2ex}
\end{align}
where $C_i$ is the $i$th extracted community, and $\frac{|C_i|}{|V|}$ is its relative size w.r.t. the whole graph. Then we pick the $\tau_1$ that maximizes the information entropy, which is
\begin{align}\vspace{-2ex}
    \tau_1 = \argmax_{\tau_1} -\sum_i \frac{|C_i|}{|V|} \cdot \log \frac{|C_i|}{|V|}. \label{eq:tau1}\vspace{-2ex}
\end{align}

The second principle is, no isolated vertex. We believe that in real-world networks, especially in social networks, each node (user) should be included in some communities as long as they are connected to some neighbors, and pure isolated vertices do not exist. According to this, $\tau_2$ should satisfy:
\begin{align*}\vspace{-2ex}
    \tau_2 \leq \min_i \max_j w_{ij}.\vspace{-2ex}
\end{align*}
To prevent assigning two vertices with very low similarity to the same community, we set
\begin{align}\vspace{-2ex}
    \tau_2 = \min_i \max_j w_{ij}. \label{eq:tau2}\vspace{-2ex}
\end{align}

During the post-processing, we calculate $w_{ij}$ for each edge, and thresholding is also applied to every edge. These two processes each takes one round and $O(|E|)$ communication cost. Finding connected component can be done in $O(\log d)$ rounds with $O(|V| + |E|)$ communication cost per round, where $d$ is the diameter of the graph \cite{Chitnis:2013:FCC:2510649.2511220}. $\tau_2$ can be calculated in one round with $O(|V|)$ communication cost. And $\tau_1$ can be found by enumerating possible values within $[\tau_2, \max w_{ij}]$ with small intervals (usually $0.001$ is enough), which takes constant times of thresholding and finding connected components. Thus, the whole post-processing takes $O(\log d)$ rounds, each with $O(|E| + |V|)$ communication cost.



\section{Incremental Updating}
\label{section_incremental}

In this section we discuss how to incrementally update the current results when the graph is changed, instead of running the whole algorithm again. Latter on an incremental algorithm is proposed and the efficiency is analyzed.

Before we start the discussion, some premises should be clarified. First of all, Algorithm \ref{alg:rslpa} contains a post-processing to the label propagation result, which drops much of the label information, but in order to update the result accurately, we need the complete result. Therefore, our strategy is to incrementally update the label sequences after label propagation, and then use the post-processing to extract communities. Second, for the changes on the graph, we will focus on edge insertion and deletion. Vertex insertion can be handled in the same way as pretending the new vertex was an old vertex with all old neighbors removed. Vertex deletion can also be handled by ignoring the deleted vertex. Furthermore, we assume the inserted and deleted edges are both picked uniformly, so each existing edge will have equal probability to be deleted, and each non-existing edge will have equal probability to be inserted.

\vspace{-1ex}
\subsection{Handling Adjacent Edge Changes}
\label{subsec_update_strategy}

After $T$ times of label propagation, there are $T+1$ labels in the label sequence of each vertex. We use $l_i^k$ to denote the $k$th label in $L_i$, where $l_i^0=i$ is the initial label of $v_i$, and $l_i^t, t>0$ is the label that $v_i$ picked at iteration $t$. We also use $L_i^t$ to denote the label sequence of $v_i$ after $t$ iterations, i.e., $L_i^t=(l_i^0, l_i^1, \ldots, l_i^t)$.


When an edge is inserted or deleted, labels of its adjacent vertices might be influenced. According to Algorithm \ref{alg:rslpa}, the source vertex of each label should be picked uniformly from the neighbor set of each vertex. When the graph changes, some of the sources cannot be treated as uniformly picked from the neighbor sets any more. This is because the neighbor sets of some vertices are changed and need to be carefully handled. On the other hand, as long as the previous results can still be treated as uniformly picked from neighbor set on the new graph, we can keep those results unchanged to save computation resources. One way to understand this is to pretend that we use the same series of random numbers to perform label propagation on the new graph.

Based on how the neighbor set of a vertex changes, we classify the vertices into 3 categories. For each of those categories, we carefully analyze whether a source can be kept or not, and how to deal with it so that we can keep as many results as possible. This will be the fundamental of the incremental algorithm.


\textbf{Category 1: Vertices with no neighbor changed.} For such a vertex $v_i$, since the source $src_i^t$ was uniformly picked from the same neighbor set on the old graph, we can keep $src_i^t$ and $pos_i^t$ unchanged.

\begin{figure}[htbp]
    \centering\vspace{-3ex}
    \subfloat[A label $3$ is picked through a preserved edge. This label can be kept unchanged.]{
        \includegraphics[width=0.3\linewidth, keepaspectratio]{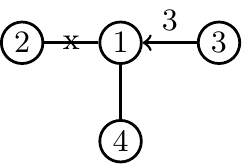}
    }%
    \hspace{0.15\linewidth}
    \subfloat[A label $3$ is picked through a deleted edge. A new label should be picked from current neighbors.]{
        \includegraphics[width=0.3\linewidth, keepaspectratio]{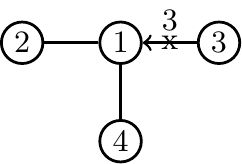}
    }\vspace{-1ex}
    \caption{Two situations in Category 2. Edges with crosses ``X'' are deleted edges. Edges with arrows and labels above mean that one label is picked through this edge.}
    \label{fig:category2}\vspace{-3ex}
\end{figure}

\textbf{Category 2: Vertices that only lose neighbor(s).} For such a vertex $v_i$, we need to check if the source $src_i^t$ of the current label is among the removed neighbor(s), or in other words, if the label is picked through a deleted edge. If so, it means the chosen edge no longer exists, so we have to pick a new label. If not, we can keep the label unchanged, because the source can still be treated as uniformly picked from the remaining neighbors. Figure \ref{fig:category2} shows examples for those two situations. We show that the result of the second situation can be treated as uniformly picked through all current edges in Theorem \ref{theorem:1}.

\begin{theorem}
    \label{theorem:1}
    In Category 2, a label comes from a remaining neighbor can still be treated as uniformly picked from all remaining neighbors.
    \begin{proof}
        Let $N=\{v_1, v_2, \ldots, v_n\}$ be the neighbor set of a vertex, of which one is picked uniformly as $s$, i.e.,
        \begin{align*}
            \mathbb{P}(s=v)=\frac{1}{n}, \forall v \in N.
        \end{align*}
        Then $n-k$ uniformly picked neighbors are removed from $N$. The rest neighbors are denoted as $N'=\{v'_1, v'_2, \ldots, v'_k\}$, so
        \begin{align*}
            \mathbb{P}(v \in N')=\frac{k}{n}, \forall v \in N.
        \end{align*}
        We also know that $s$ is in $N'$, which can be written as
        \begin{align*}
            \mathbb{P}(s \in N')=1.
        \end{align*}
        Then the probability that a neighbor in $N'$ is equal to $s$ is
        \begin{align*}
            \mathbb{P}(s=v|v \in N')&=\frac{\mathbb{P}(s=v \land v \in N')}{\mathbb{P}(v \in N')} \\
                                    &=\frac{\mathbb{P}(s=v \land s \in N')}{\mathbb{P}(v \in N')} \\
                                    &=\frac{\mathbb{P}(s=v)}{\mathbb{P}(v \in N')}=\frac{1}{k}.
        \end{align*}
    \end{proof}
\end{theorem}

\begin{figure}[htbp]
    \centering\vspace{-3ex}
    \subfloat[A label $3$ is picked through a preserved edge. With probability 1/3, a new label should be picked from vertex 4.]{
        \includegraphics[width=0.3\linewidth, keepaspectratio]{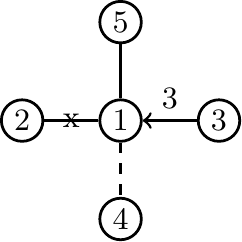}
    }%
    \hspace{0.15\linewidth}
    \subfloat[A label $3$ is picked through a deleted edge. A new label should be picked from current neighbors.]{
        \includegraphics[width=0.3\linewidth, keepaspectratio]{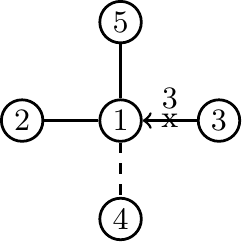}
    }\vspace{-1ex}
    \caption{Two situations in Category 3. Edges with a cross are deleted edges. Dashed edges are inserted edges. Edges with an arrow and a label above mean that one label is picked through this edge.}
    \label{fig:category3}\vspace{-3ex}
\end{figure}

\textbf{Category 3: Vertices with new neighbor(s).} For such a vertex $v_i$, there are new edges that we are not aware of when we were performing label propagation on the old graph, in addition, there might be edges that no longer exist. According to Theorem \ref{theorem:1}, if the source of $l_i^t$ remains to be one of the neighbors, it can be treated as uniformly picked from unchanged neighbors on the new graph. To extend this to all current neighbors, we can add another random process. Suppose the number of unchanged neighbors is $n_u$ and the number of new neighbors is $n_a$. Then, with probability $\frac{n_u}{n_u + n_a}$, we keep the label unchanged, otherwise we uniformly pick a label from the new neighbors. After this random process, the current label $l_i^t$ is uniformly picked from all current neighbors. The proof is shown in Theorem \ref{theorem:2}. If the source of $l_i^t$ is deleted, we have to pick a label from all current neighbors. Figure \ref{fig:category3} shows examples for this category.

\begin{theorem}
    \label{theorem:2}
    For a label coming from an unchanged neighbor, after the random process in Category 3, the result label is uniformly picked from all current neighbors.
    \begin{proof}
        Suppose the original, unchanged and added neighbor set are $N$, $N_u$ and $N_a$, each of size $n$, $k$ and $m$, respectively. The current neighbor set $N_c$ can be represented as $N_u \cup N_a$, which is of size $k+m$. $s$ is picked uniformly from $N$, and $s$ is in $N_u$.

        After the random process, if the source of the result $r$ equals to $s$ with probability $\frac{k}{k+m}$, then integrating the result from Theorem \ref{theorem:1}, we have
        \begin{align*}
            \mathbb{P}(r=v|v \in N_u)&=\frac{k}{k+m}\cdot\mathbb{P}(s=v|v \in N_u)=\frac{1}{k+m}.
        \end{align*}

        If $r$ is picked from $N_a$ with probability $\frac{m}{k+m}$, then
        \begin{align*}
            \mathbb{P}(r=v|v \in N_a)&=\frac{m}{k+m}\cdot\frac{1}{m}=\frac{1}{k+m}.
        \end{align*}

        Combining those two, we have
        \begin{align*}
            \mathbb{P}(r=v|v \in N_c)=\frac{1}{k+m}.
        \end{align*}
    \end{proof}
\end{theorem}

\vspace{-1ex}
\subsection{Handling Subsequent Updates}
\label{subsec_subsequential_updates}

Besides the label updates that are caused by graph changes directly, there is another kind of label updates that we should take care of. Since each label $l_i^t$ is picked from another label $l_j^k$, if $l_j^k$ is updated, then $l_i^t$ also needs to be updated correspondingly, if $l_i^t$ keeps $v_j$ as its source. Moreover, $l_i^t$ can be picked in later iterations and thus may cause the updates of other labels. Actually, a specific label can be propagated through several edges in different iterations, which form a propagation path or even propagation tree. Once an edge of the tree is changed, all subsequent labels need to be updated accordingly. Example \ref{exa:sub_change} shows a situation where the deletion of one edge causes updates on three more labels.

\begin{figure}[htbp]
    \centering\vspace{-3ex}
    \includegraphics[width=0.4\linewidth, keepaspectratio]{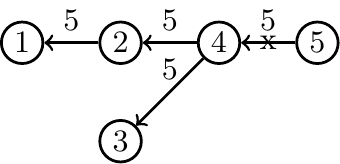}\vspace{-1ex}
    \caption{A possible propagation tree of label $5$, with one edge deleted.}
    \label{fig:propagation_tree}\vspace{-3ex}
\end{figure}

\begin{example}
\label{exa:sub_change}
In Figure \ref{fig:propagation_tree} there is a possible propagation tree of a label, where the edge between vertices $4$ and $5$ is the first edge. If this edge is deleted as the figure shows, then vertex $4$ has to pick a new label to replace the label $5$ picked from vertex $5$. Thereafter, the label $5$ of vertices $3$, $2$ and $1$ all need to be updated to the newly-picked label, if no other changes are made.
\end{example}

To handle this kind of updates, we need to record $src_i^t$ and $pos_i^t$ for each picked label $l_i^t$, so that the change of source label can be detected by comparing the two labels. Then we can update $l_i^t$ if necessary.

Now we can come up with a simple version of the updating process: We enumerate the iteration number $t$ from 1 to $T$, then check and update the existing result of iteration $t$, and finally obtain the result that satisfies the required probability distribution on the new graph. However, this version requires $O(T)$ rounds, each with communication cost $O(|V|)$, which is no better than re-running the whole algorithm on the new graph. The reason is that, no matter what the situation is, we need to retrieve a label from one neighbor to update or check the current label. This operation is equivalent to picking a random label from a random neighbor, as we did in the original algorithm.

We believe that most of the labels will remain unchanged. When $src_i^t$ and $pos_i^t$ are both unchanged, there is a high chance that the current label remains the same. Thus, instead of checking the correctness of each label aggressively, we wait until the neighbor informs the change of labels. Following this idea, we design an internal data structure to enable the forwarding of label change information. The data structure, as a reverse record of $src_i^t$ and $pos_i^t$, records the receivers of each label in the label sequence. Let $R_i$ denote the records of $v_i$, that is $R_i=\{R_i^0, R_i^1, \ldots, R_i^T\}$, where $R_i^t$ is the set of pairs $(tar, k)$ denoting that neighbor $tar$ picked $l_i^t$ in iteration $k$. The whole set of records is then denoted by $R=\{R_0, R_1, \ldots, R_n\}$. $R_i$ can be simply recorded during the label propagation process with no additional operations required.



With $R_i$ computed, once a label $l_i^t$ is changed, $v_i$ needs to send the change information to the neighbors in $R_i^t$; those neighbors can update their labels with $src=i$ and $p=t$, and then send the change information if necessary. As each label $l_i^t$ can only be picked by other vertices in the later iterations, any change information can only trigger consequent change information with bigger iteration number. Thus, all the label change can be handled within $T$ iterations.

When the graph changes, $R_i$ itself may change and need the maintenance. When $v_i$ decides to switch from $src_i^t$ and $pos_i^t$ to ${src'}_i^t$ and ${p'}_i^t$, and there are no other labels with the same $src_i^t$ and $pos_i^t$, one additional information should be sent to the original source $src_i^t$, so that it can remove $v_i$ from the corresponding receiver list(s). When $v_i$ fetches the new label from ${src'}_i^t$, it will be added to the receiver lists of the new source. By amortized analysis, these additional operations will not increase the complexity of the algorithm.

\vspace{-1ex}
\subsection{Correction Propagation Algorithm}

Combining the two kinds of updates we discussed above, we can now write the complete algorithm to incrementally update the label sequence of each vertex. The main steps are summarized in Algorithm \ref{alg:inc_prop}. First, we check all vertices for adjacent edge changes. A function named ``NeedRepick'' is used to check which of the three categories a vertex belong to, and another function named ``Repick'' is used to pick a new $src$ and $pos$ according to different situations as discussed before. From line 1 to line 12, we handle the adjacent edge changes and send label requests to neighbors. The $sendbuffer$ is used to store the information to send to neighbors temporarily. In lines 14 and 15, we send the labels to receivers, and in lines 16 to 23, we update the label sequences and then compute the correction information that needs to be sent to neighbors. When all vertices have no information to send to neighbors, it means all updates are done, and we stop.


\begin{algorithm}
    \caption{Correction Propagation}
    \label{alg:inc_prop}
    \SetKwInOut{Input}{Input}\SetKwInOut{Output}{Output}
    \SetKwFunction{NeedRepick}{NeedRepick}
    \SetKwFunction{Repick}{Repick}
    \SetKwFunction{Push}{Push}
    \SetKwFunction{Pop}{Pop}
    \SetKwFunction{NotEmpty}{NotEmpty}
    \SetKwFunction{Update}{Update}
    \SetKwProg{Mapper}{Mapper}{:}{}
    \SetKwProg{Reducer}{Reducer}{:}{}
    \Input{$L$ produced by Algorithm \ref{alg:rslpa}, additional records $R$ and the changed graph $G'(V', E')$.}
    \Output{Maintained $L$ and $R$ to fit the new graph.}
    \BlankLine
    \tcc{Handle adjacent edge changes.}
    \Mapper{for $v_i$}{
        \For{$\; t \leftarrow 1$ \KwTo $T \;$}{
            \If{\NeedRepick{$label_i^t$}}{
                $src, pos \leftarrow \Repick{t}$\;
                Emit $(src, pos, i, t)$\;
            }
        }
    }
    \Reducer{for $v_i$}{
        \ForEach{received $\; (pos, j, t) \;$}{
            Update $R_i$\;
            Add $(j, l_i^{pos}, t)$ to $buffer_i$
        }
    }
    \tcc{Receive updated labels and perform correction propagation.}
    \While{ Any $buffer_i$ is not empty}{
        \Mapper{for $v_i$}{
            Emit each $(j, l, t)$ in $buffer_i$\;
        }
        \Reducer{for $v_i$}{
            $buffer_i \leftarrow \emptyset$\;
            \ForEach{received $\; (l, t) \;$}{
                $l_i^t \leftarrow l$\;
                \ForEach{$\; (tar, k) \in R_i^t \;$}{
                    Add $(tar, l_i^t, k)$ to $buffer_i$\;
                }
            }
        }
    }
\end{algorithm}

The benefit is that we only visit vertices that are close to the changed edges. Labels that need no updates will not appear in the emitted information and thus are naturally filtered out.

\vspace{-1ex}
\subsection{Complexity of Correction Propagation Algorithm}
The Correction Propagation algorithm only updates labels with necessity, so the running time of the algorithm is decided by the number of labels that need updates. Since this number is a random variable due to the randomness of the algorithm,
we will give the estimation of this number as well as the best and worst cases rather than a fixed value.

\subsubsection{The Expected Complexity}
For simplicity, we use $\eta$ to denote the number of labels that need updates. One way to calculate $\eta$ is to first estimate how many labels are needed to update because of one single edge insertion or deletion, and then multiply this number by the total number of the changed edges. However, multiple edges can lie on the same path, making it very difficult to generalize the situation of a single edge to multiple edges. Instead, to estimate the probability that a single label needs to be updated is a more promising way. Let $P$ denote this probability, and then $\eta=P \cdot T \cdot |V|$, where $T \cdot |V|$ is the total number of labels picked after $T$ iterations.

Directly estimating $P$ is still a hard job, since labels picked in different iterations have correlations. For example, suppose label $l_i^t$ is picked from $l_j^k$. Then once $l_j^k$ needs to be updated, $l_i^t$ also needs to be updated accordingly. To make those correlations clearer, we will estimate $P(t)$, which is the expected probability that a label picked in iteration $t$ needs to be updated. Moreover, we found that $Q(t)=1-P(t)$, which is the probability of the complementary event, has a much simpler form, so we will focus on estimating $Q(t)$, where $P(t)$ is given by $1-Q(t)$.

We start from a special case. When $t=0$, $Q(0)$ is the probability that initial label of each vertex will not change, which always holds based on our algorithm. Thus, $Q(0)=1$.

When $t=1$, all labels are picked from the initial label of each
vertex, which will never change. Thus the label will need update if
(1) the chosen edge is deleted, or (2) the chosen edge is not
deleted but switched to a new edge. Assuming that edges are deleted
and inserted randomly with no prior distribution, then probability of Condition 
(1) is $\frac{m_d}{|E|}$, where $m_d$ is the number of deleted
edges. The probability of Condition (2) is a bit more complex. According to the
discussion of Category 3 in Section \ref{subsec_update_strategy}, the
probability of Condition (2) can be written as
$(1-\frac{m_d}{|E|})(\frac{n_u}{n_u+n_a})$. Similarly, with no
priori knowledge about the distribution of vertex degrees, $n_u$ is
estimated by $\frac{|E|-m_d}{|V|}$, and $n_a$ is estimated by
$\frac{m_a}{|V|}$, where $m_a$ is the number of inserted edges. Now
the probability of Condition (2) is written as
$(1-\frac{m_d}{|E|})(\frac{|E|-m_d}{|E|-m_d+m_a})$. Integrating Conditions (1)
and (2), the probability that a chosen edge is changed is:\vspace{-3ex}

\begin{align}
    p_c&=\frac{m_d}{|E|}+(1-\frac{m_d}{|E|})(\frac{|E|-m_d}{|E|-m_d+m_a}),
\end{align}
and using $p_c$ we can write $P(1)$ and $Q(1)$ as:
\begin{align}
    P(1)&=p_c \cdot Q(0)+1-Q(0) = p_c, \\
    Q(1)&=1-p_c. \label{equation_q1}
\end{align}\vspace{-3ex}

When $t>1$, each label can be chosen from any labels picked in the previous iterations ($t$ choices in total), which causes the correlation between labels. Fortunately, we can utilize this relationship to simplify the derivation. For each of the $t$ choices, the $k$th label is picked ($1 \leq k \leq t$) with probability $\frac{1}{t}$, and it needs no update as long as the following two conditions hold: (1) with probability $Q(k)$, the label itself does not change, and (2) with probability $Q(1)$, the chosen edge is not deleted nor switched to another one. Combining these two conditions, we now have a recursion formula:
\begin{align}
    Q(t)&=\sum_{k=0}^{t-1}\frac{1}{t}(1-p_c)Q(k) \nonumber \\
        &=\frac{t-1}{t}Q(t-1)+\frac{1}{t}(1-p_c)Q(t-1) \nonumber \\
        &=(1-\frac{p_c}{t})Q(t-1). \label{equation_qt}
\end{align}

With the first term $Q(1)$ from Equation \ref{equation_q1} and the recursion formula from Equation \ref{equation_qt}, we are now able to write the general term of $Q(t)$:
\begin{align}
    Q(t)=\prod_{k=1}^{t}(1-\frac{p_c}{k}).
\end{align}
Note that the first term $Q(1)$ can also be represented by this general form.

Finally, we can write the estimation of $\eta$ as:
\begin{align}
    \hat\eta&=\sum_{t=1}^{T}|V| \cdot P(t) \nonumber \\
        &=\sum_{t=1}^{T}|V| \cdot (1-Q(t)) \nonumber \\
        &=T \cdot |V|-|V| \cdot \sum_{t=1}^{T}Q(t) \nonumber \\
        &=T \cdot |V|-|V| \cdot \sum_{t=1}^{T}\prod_{k=1}^{t}(1-\frac{p_c}{k}). \label{equation:ita}
\end{align}
The expected running time of the centralized incremental algorithm is $O(\hat\eta)$. For distributed version, the overall communication cost is $O(\hat\eta)$.

Besides this expectation, we would like to investigate the upper and lower bounds of the running time to better describe how the actual running time varies among different instances. Later we will discuss these two bounds under the assumption that no prior distribution is given about the degrees of vertices and the choices of editing edges.

\subsubsection{The Best Case}

Recall that $Q(t)$ is derived from Equation \ref{equation_qt}. From this equation we can further derive the upper bound of $Q(t)$:
\begin{align}
    Q(t)&=(1-\frac{p_c}{t})Q(t-1) \leq Q(t-1), \nonumber
\end{align}
and by conductivity,
\begin{align}
    Q(t)\leq Q(1)=1-p_c.
\end{align}
Then the number of labels that need the update has a lower bound as
\begin{align}
    \hat\eta=|V|\sum_{t=1}^{T}(1-Q(t)) \geq T|V|\cdot p_c.
\end{align}
This lower bound refers to the situation where every label is picked from the initial labels of the chosen neighbor. Consequently all paths of the picked labels have length 1, which is the minimum possible length, so the number of labels influenced by each edit edge is minimized, which gives the best running time.

\subsubsection{The Worst Case}

Similarly, the lower bound of $Q(t)$ can also be derived from Equation \ref{equation_qt}:
\begin{align}
    Q(t)&=(1-\frac{p_c}{t})Q(t-1) \nonumber \\
        &\geq (1-p_c)Q(t-1) \nonumber \\
        &=(1-p_c)^t.
\end{align}
Then the $\eta$ has an upper bound as
\begin{align}
    \hat\eta&=T \cdot |V|-|V| \cdot \sum_{t=1}^{T}Q(t) \nonumber \\
        &\leq T \cdot |V|-|V| \cdot\sum_{t=1}^{T}(1-p_c)^t \nonumber \\
        &=T \cdot |V|-|V| \cdot \frac{1-p_c-(1-p_c)^{T+1}}{p_c}.
\end{align}
This refers to the situation where every label in iteration $t$ is picked from labels of iteration $t-1$. Thus paths of the labels picked at iteration $t$ have length $t$, which is the maximum possible length. So the number of labels influenced by each edit edge is maximized, and this gives the worst running time.

\vspace{-1ex}
\section{Experiments}
\label{section_exp}
To evaluate the effectiveness and efficiency of rSLPA, we conduct extensive experiments on both synthetic and real-world graphs. We use generated networks that contain known communities to test the quality of communities given by rSLPA. Then we use large-scale real-world networks to test its efficiency in both static and dynamic scenarios. We compare our rSLPA with SLPA on both effectiveness and efficiency.

\vspace{-1ex}
\subsection{Synthetic Dataset}

\begin{figure*}[t]
    \centering\vspace{-4ex}
    \subfloat[NMI score with different iterations.]{
        \includegraphics[width=.28\linewidth, keepaspectratio]{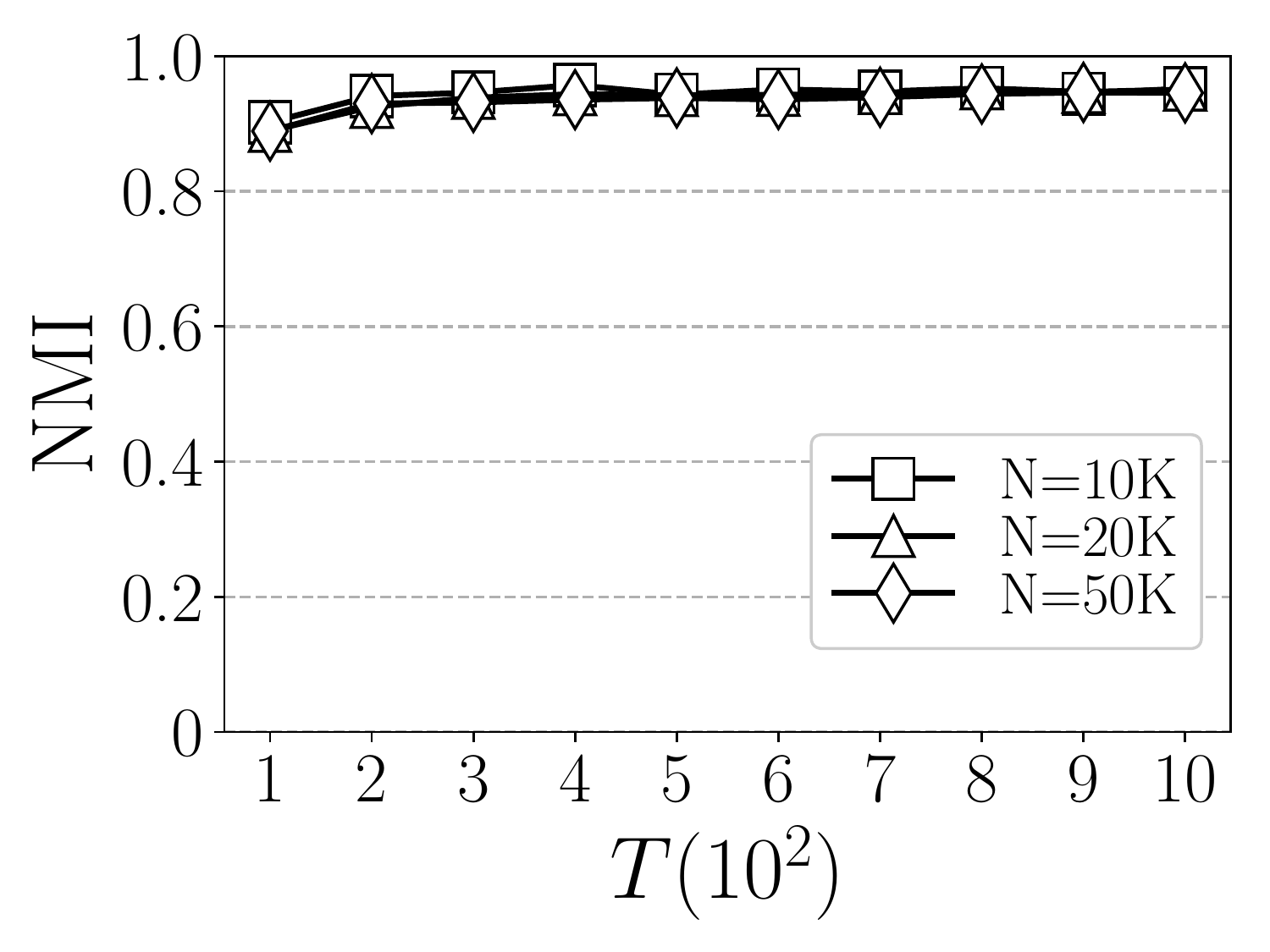}
        \label{fig:exp_converge}
    }%
    \hspace{0.02\linewidth}
    \subfloat[NMI score when varying $N$.]{
        \includegraphics[width=.28\linewidth, keepaspectratio]{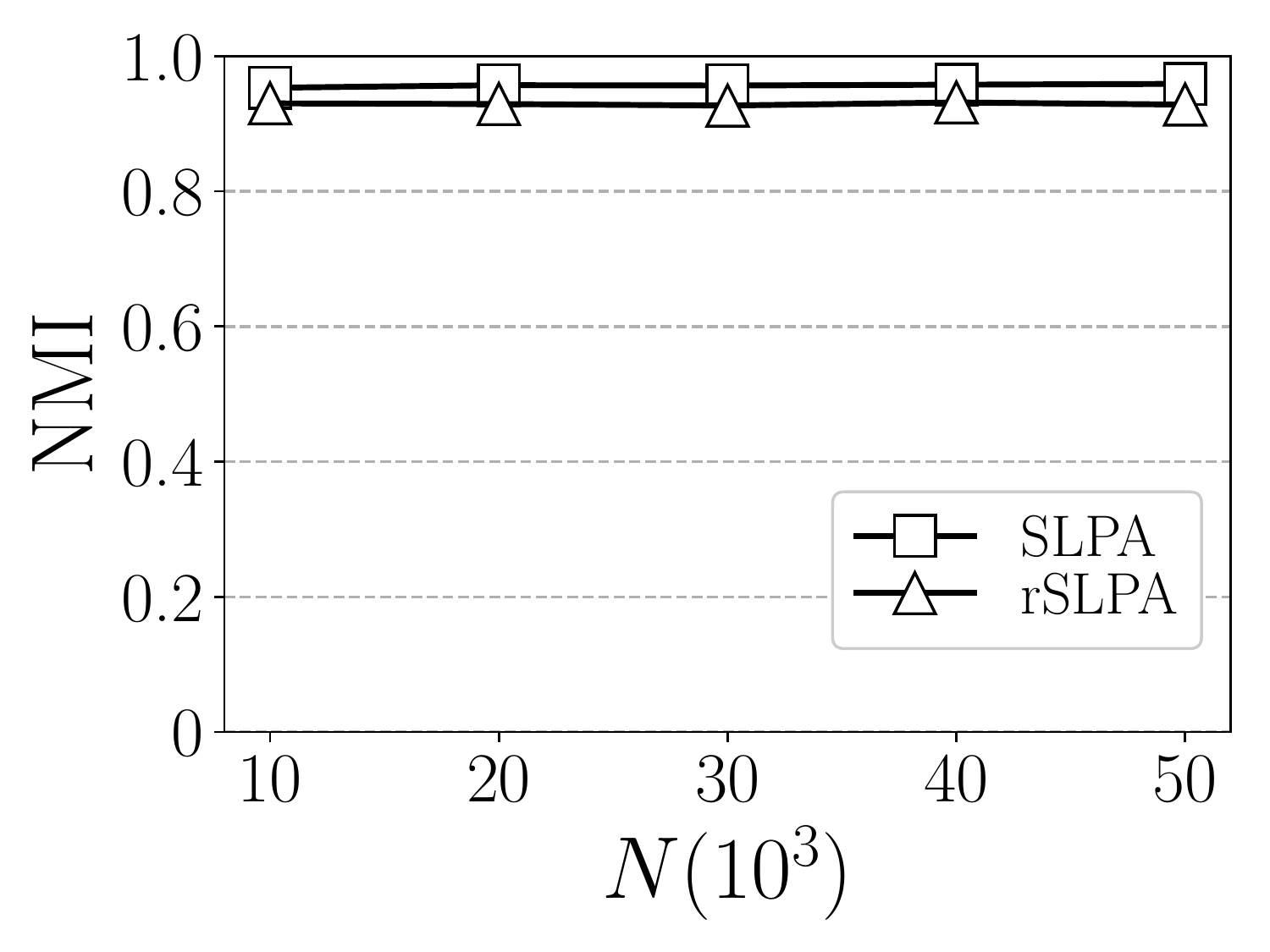}
        \label{fig:exp_quality_N}
    }%
    \hspace{0.02\linewidth}
    \subfloat[NMI score when varying $k$.]{
        \includegraphics[width=.28\linewidth, keepaspectratio]{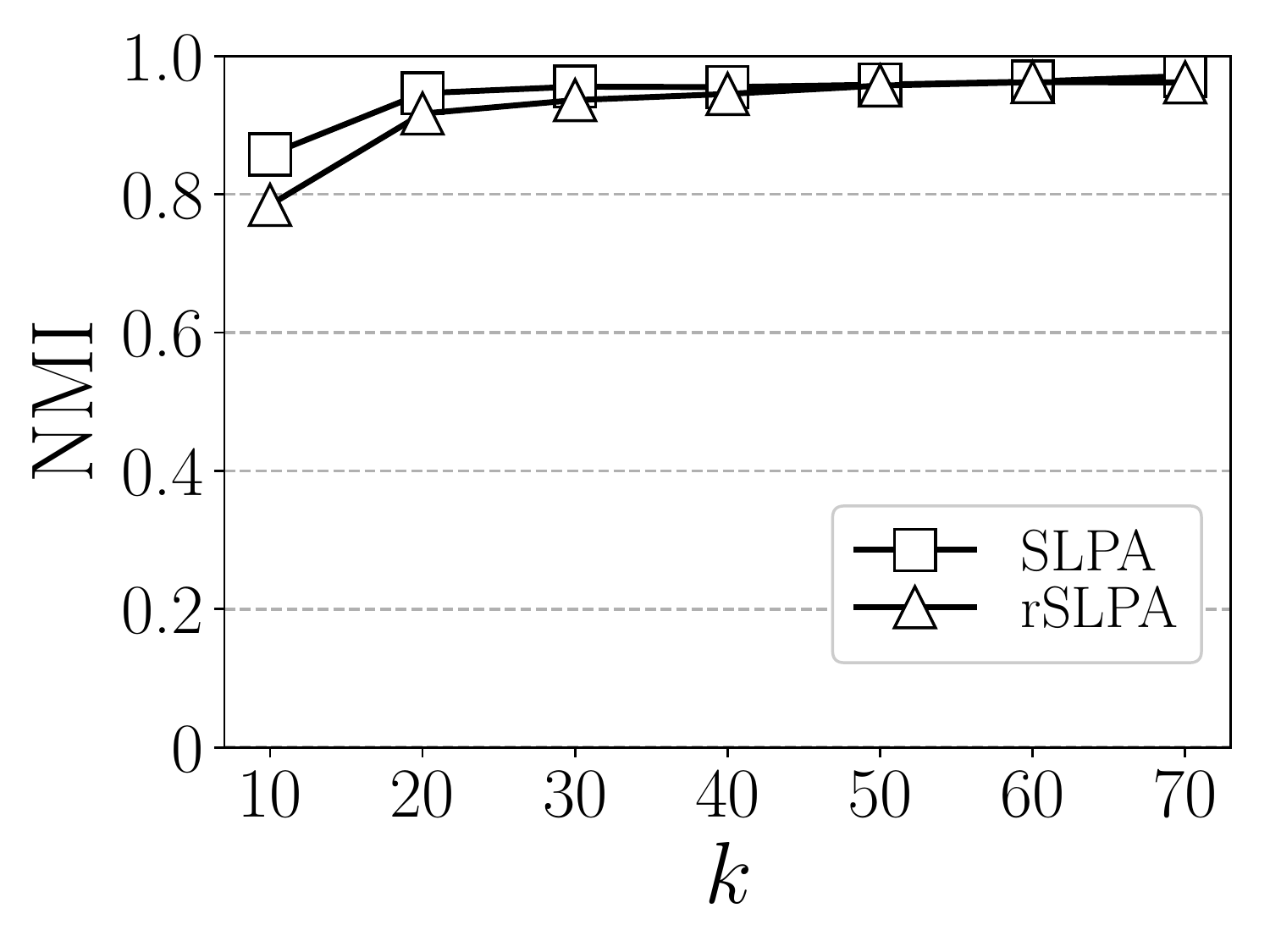}
        \label{fig:exp_quality_k}
    }%
    \hspace{0.02\linewidth}
    \subfloat[NMI score when varying $\mu$.]{
        \includegraphics[width=.28\linewidth, keepaspectratio]{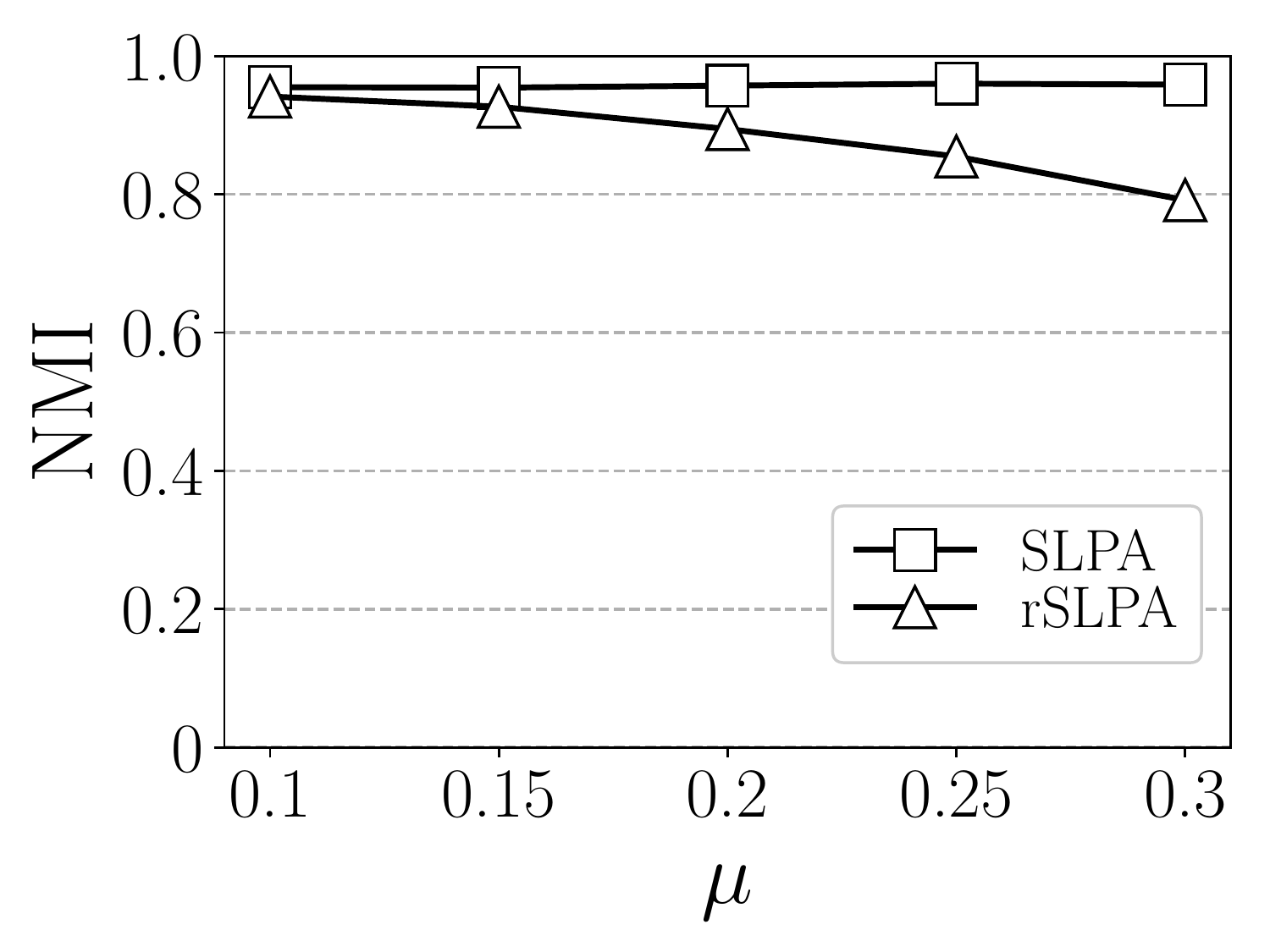}
        \label{fig:exp_quality_mu}
    }%
    \hspace{0.02\linewidth}
    \subfloat[NMI score when varying $om$.]{
        \includegraphics[width=.28\linewidth, keepaspectratio]{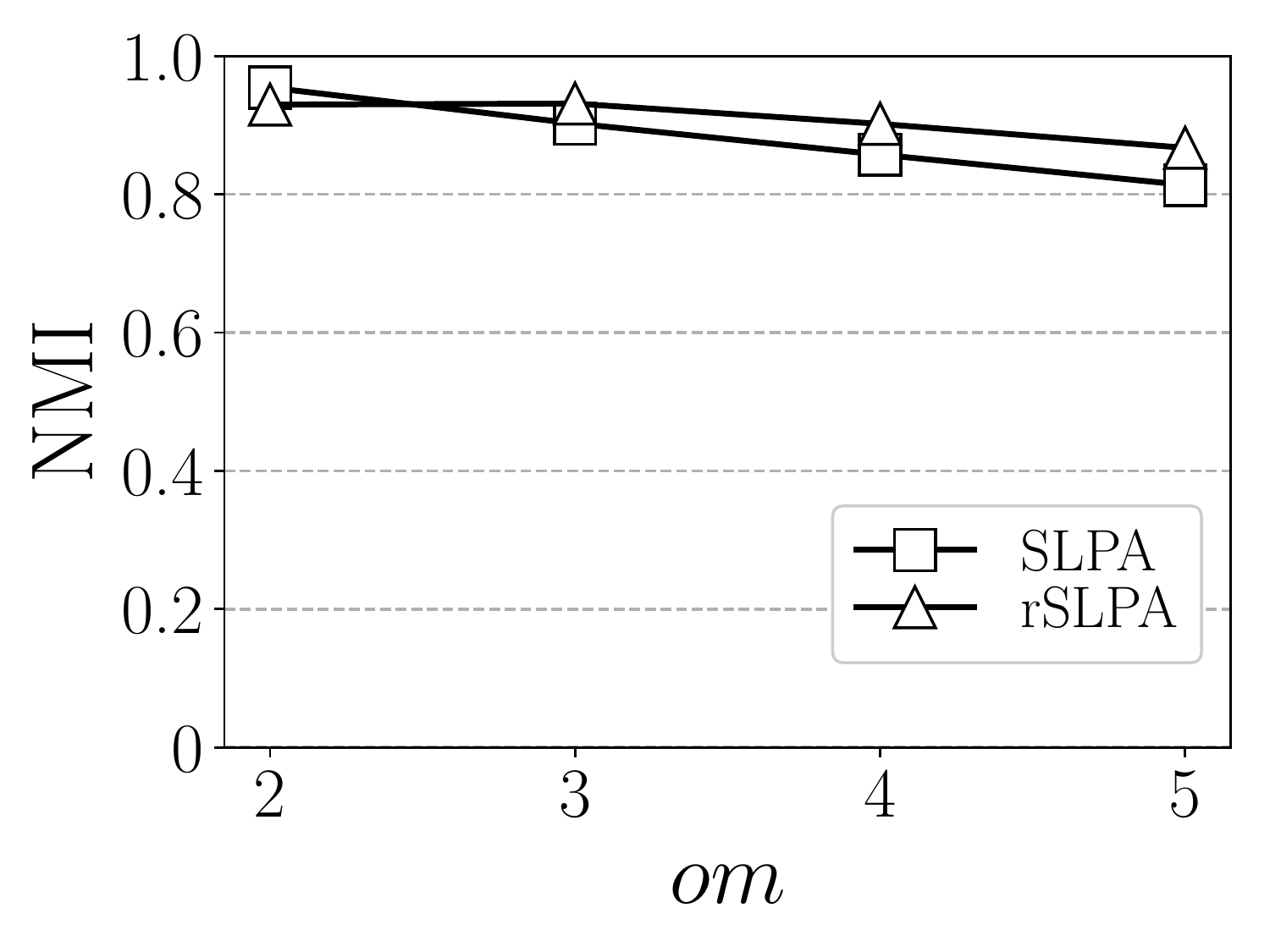}
        \label{fig:exp_quality_om}
    }%
    \hspace{0.02\linewidth}
    \subfloat[NMI score when varying $on$.]{
        \includegraphics[width=.28\linewidth, keepaspectratio]{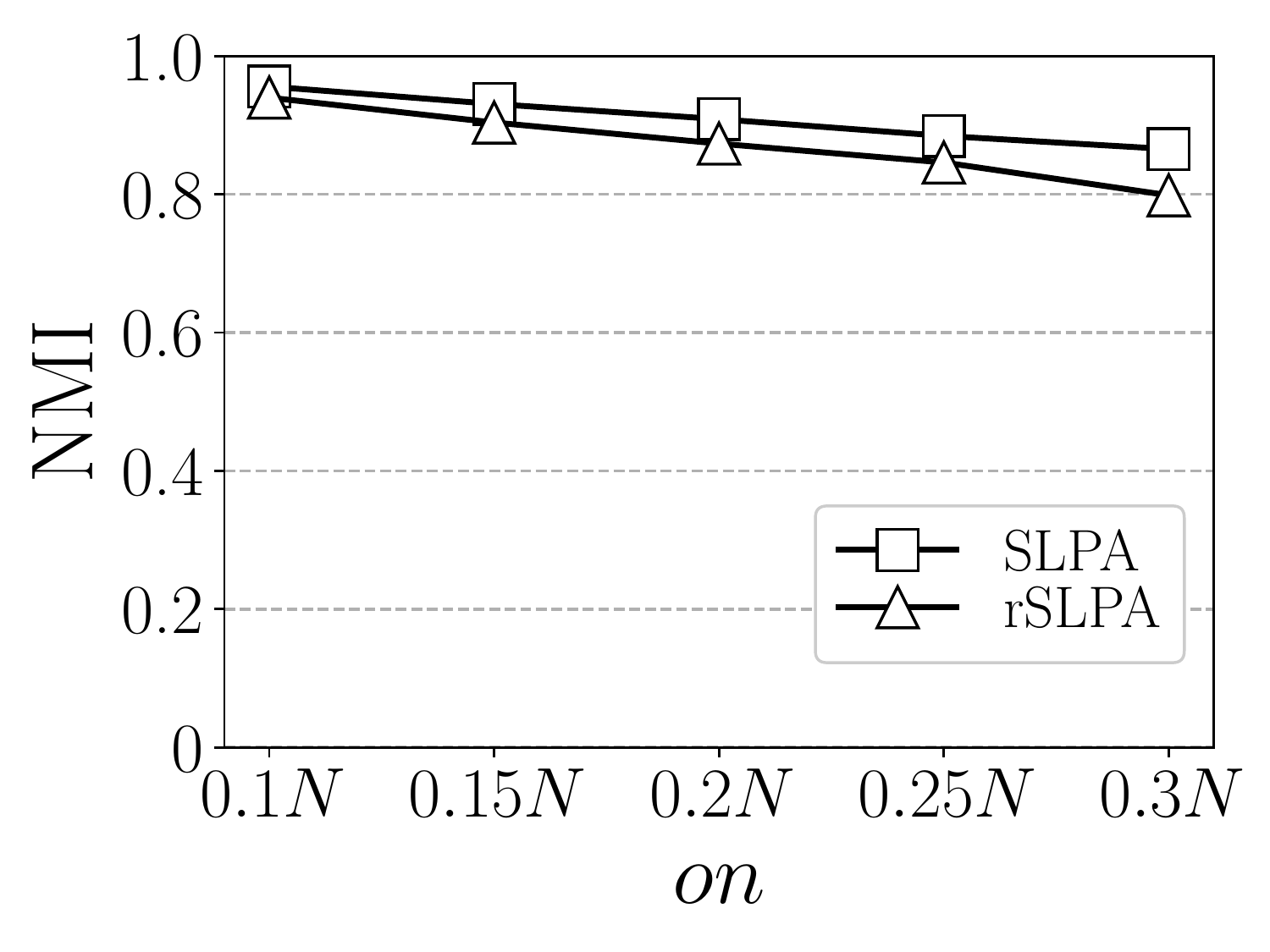}
        \label{fig:exp_quality_on}
    }
    \caption{Experiment results on synthetic dataset.}\vspace{-3ex}
\end{figure*}

\subsubsection{Data Generation}
We use the LFR benchmark \cite{lancichinetti2009benchmarks} to generate graphs with known communities. This benchmark overcomes the drawbacks of the GN benchmark \cite{girvan2002community} and is widely used in evaluating community detection algorithms. The most important parameters to generate a graph are listed in Table \ref{table:lfr_param}. We choose $N=10,000$, $k=30$, $maxk=100$, $om=2$, $on=0.1N$ and $\mu=0.1$ as the default setting and keep the value of each parameter unchanged unless specified.

\begin{table}
    \centering\vspace{-2ex}
    \caption{Parameters of the LFR Benchmark (Selected)}
    \begin{tabular}{|c|c|} \hline
        \textbf{Parameter}  & \textbf{Description} \\ \hline
        $N$     & the number of vertices \\ \hline
        $maxk$  & the max degree \\ \hline
        $k$     & the average degree \\ \hline
        $\mu$ & the mixing parameter \\ \hline
        $on$ & the number of overlapping vertices \\ \hline
        $om$ & the number of memberships of overlapping vertices \\ \hline
    \end{tabular}\vspace{-3ex}
    \label{table:lfr_param}
\end{table}

\subsubsection{Evaluation Metrics}
In community detection, with known communities as the ground truth, the \textit{Normalized Mutual Information (NMI)} is one of the most widely used measures to evaluate the quality of detected communities. Basically it reveals the similarity between two membership assignments based on the information theory. The score is in range $[0,1]$, and higher value indicates higher similarity (better quality).

When a graph is generated, the LFR benchmark will also give the known communities, and we will calculate the NMI scores for each algorithm. Since both algorithms have some randomness as well as the thresholding parameters which can influence the final results, we average the NMI score over 10 runs for each single experiment.

For SLPA, \cite{xie2012towards} suggests that the iteration $T=100$ and $\tau \approx 1/om$. In our experiments, we set $T=100$ and $\tau=0.2$. For rSLPA, we set $\tau_1$ and $\tau_2$ according to Equations \ref{eq:tau1} and \ref{eq:tau2}, and we have conducted some pivot experiments for choosing the iteration number of rSLPA.

\subsubsection{Convergence Speed}
For rSLPA, we vary $T$ from $100$ to $1,000$ to test after how many iterations, the result will converge. Considering that the graph size $N$ may also affect the convergence speed, we vary the graph size in [10K, 20K, 50K].

Figure \ref{fig:exp_converge} shows the convergence speed of rSLPA. We can see that for different $N$, it gives a relatively stable result when $T \geq 200$. According to this result, we will use $T=200$ for rSLPA in the rest experiments.

\subsubsection{Community Quality}
In this part we compare the detected community quality of SLPA and rSLPA under different parameter settings. We vary the value of each single parameter $N$, $k$, $\mu$, $om$, $on$, and compare the NMI scores.

In Figure \ref{fig:exp_quality_N} we vary the graph size $N$ from
$10,000$ to $50,000$. Both algorithms have very high and stable
scores, and the difference between two algorithms is small. This
shows that, our relaxing to the voting process keeps the ability of
finding good-quality communities. In Figure \ref{fig:exp_quality_k}
we vary the average degree $k$ from $10$ to $70$, which covers
sparse and dense graphs. As $k$ increases, the score of each
algorithm grows gradually, and remains unchanged when $k$ is large
enough ($\geq 50$). Both algorithms have higher scores over denser
graphs, but they can also detect high-quality communities for sparse
graphs. From the results, we conclude that both algorithms are
suitable for graphs of different sizes and densities.

In Figure \ref{fig:exp_quality_mu}, we increase the mixing parameter $\mu$ from $0.1$ to $0.3$. The score of SLPA is nearly unchanged as $\mu$ increases. On the other hand, the score of rSLPA also remains at a high level, but it drops slowly as $\mu$ increases, which means rSLPA has less ability to detect better-mixed communities. In Figure \ref{fig:exp_quality_om}, $om$ increases from $2$ to $5$, which means each overlapping vertex belongs to more and more communities simultaneously. The scores of both algorithms decrease slowly as $om$ increases. The reason is that it becomes harder to correctly assign a vertex when it belongs to more communities. Compared to SLPA, rSLPA has better performance when $om \geq 3$. This means for a single vertex, rSLPA keeps more information about its belongingness, so that rSLPA can correctly detect more memberships of a vertex. This confirms the discussion in Section \ref{subsec:uniform} from another perspective. In Figure \ref{fig:exp_quality_on}, the number of overlapping vertices is increased from $0.1N$ to $0.3N$. With the increase of overlapping vertices, the performance of both algorithms becomes worse. The reason is that the boundary between communities becomes fuzzier and harder to detect.

Overall, SLPA and rSLPA keep high scores on different situations.
When $om$ is large, rSLPA can benefit from more detailed information
it keeps, and have better performance.

\vspace{-1ex}
\subsection{Real-World Dataset}
In the sequel, we use real-world dataset to test the efficiency of
SLPA and rSLPA on Spark \cite{zaharia2010spark}, a modern
cluster-computing framework based on MapReduce, to investigate its
performance in a distributed dynamic scenario.

We use $7$ Linux servers, each with 125 GBytes of main memory and 2 CPU of Intel Xeon Processor E5-2630 v3. Both algorithms are implemented in Scala.

\subsubsection{Preparing the data set}
We use a public dataset \textit{eu-2015-tpd}, which can be
downloaded from the Laboratory for Web Algorithmics
(law.di.unimi.it). This dataset consists Web pages from private
domains in Europe countries crawled in 2015. The vertices are pages
and edges are hyper-links between edges. Statistics of this dataset
are given in Table \ref{table:eu}.

\begin{table}
    \centering\vspace{-2ex}
    \caption{STATISTICS of DATASET EU-2015-TPD.}
    \begin{tabular}{|c|c|} \hline
        \textbf{Statistics}  & \textbf{Value} \\ \hline
        \# nodes             & 6,650,532 \\ \hline
        \# edges             & 170,145,510 \\ \hline
        avg. degree          & 25.584 \\ \hline
        max in-degree        & 74,129 \\ \hline
        max out-degree       & 398,599 \\ \hline
    \end{tabular}
    \label{table:eu}\vspace{-3ex}
\end{table}

The original dataset is compressed with LLP \cite{BRSLLP} and WebGraph \cite{BoVWFI}, which cannot be directly read by Spark. To make the data proper for our experiments, we extract it into plain texts, then remove the direction of edges, as well as multiple edges and self-loops.

To test the incremental algorithm, we generate the graph edit batch by randomly selecting edges for insertion and deletion. Typically, the batch size is set from $100$ to $100,000$, and then for each size we randomly pick half edges to insert and half to delete. We run rSLPA on the original graph for $200$ iterations, then apply edge insertion and deletion to the graph, and finally run the incremental updating algorithm to get the running time.

\subsubsection{Implementation}
\cite{kuzmin2013parallel} proposes a parallelized SLPA based on MPI, which uses a message buffer to passing labels between different vertices and data partitions. In this paper, we adopted the parallelized SLPA to the MapReduce model by replacing parallelized for-loops with Map and Join operations. The internal RDD produced by Map operation plays the role of the message buffer.

For rSLPA, we implement it as described in Algorithms \ref{alg:rslpa} and \ref{alg:inc_prop}. In post-processing, we slightly change the existing algorithm of finding connected components by adding filtering on edge weights, so that we do not need to explicitly generated the new graph filtered by the two thresholds.

\subsubsection{Evaluation Metrics}
We use the actual running time (wall clock time) to measure the efficiency of the algorithms. For both algorithms, the post-processing part can be done separately from the label propagation part. For example, if we run rSLPA on a social network, we may not want to calculate the communities in every minute, instead, we can let the algorithm handling changes continuously, and calculate the communities once per hour. Thus, in this part, we compare the running time of label propagation and post-processing of each algorithm separately.

\subsubsection{Results}
\begin{figure}[htbp]
    \centering\vspace{-3ex}
    \includegraphics[width=.6\linewidth, keepaspectratio]{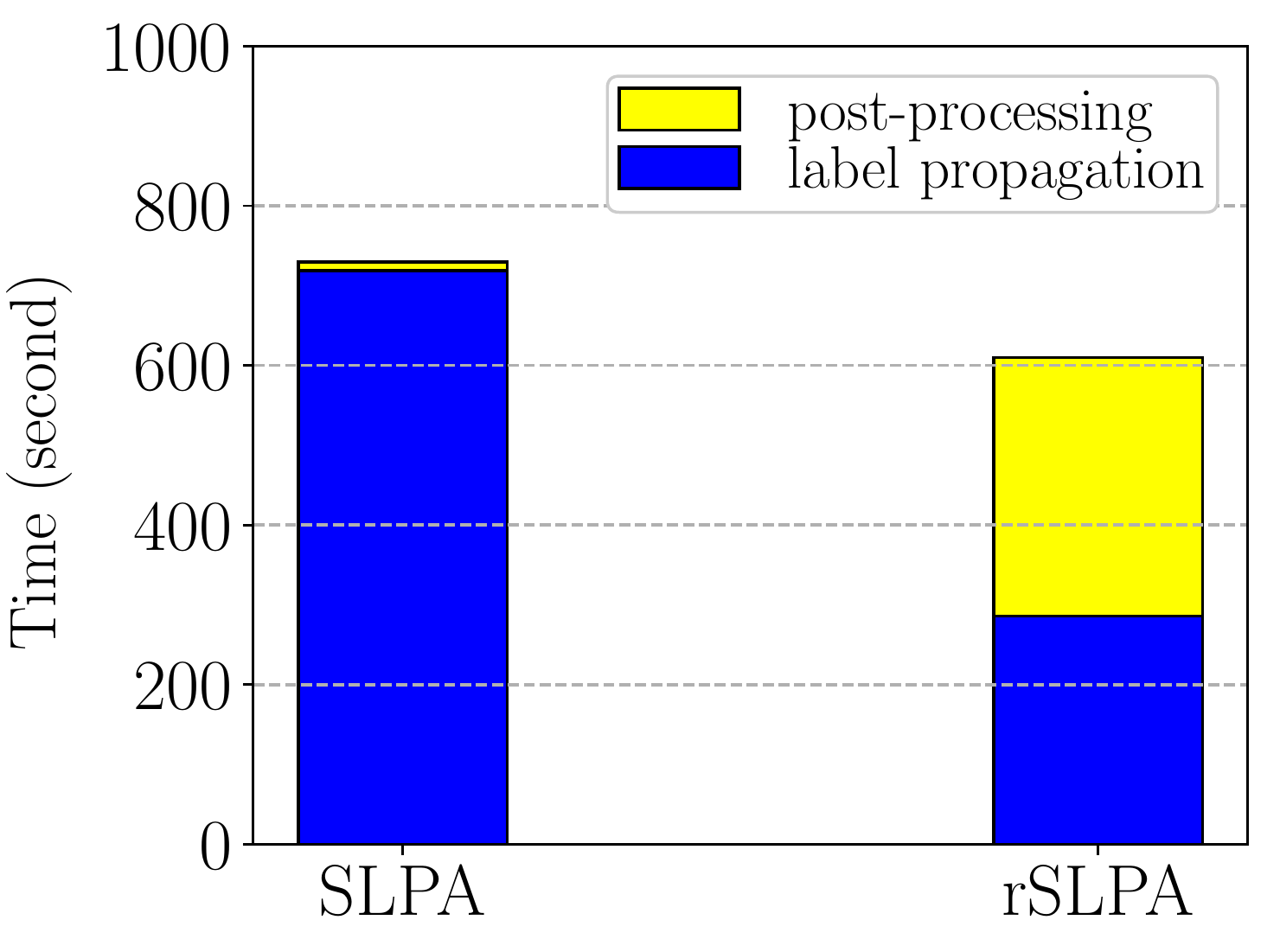}\vspace{-2ex}
    \caption{Running time of SLPA and rSLPA on real static graphs.}
    \label{fig:compare_real_static}
\end{figure}

Figure \ref{fig:compare_real_static} compares the running time of SLPA and rSLPA on real static graphs. In the label propagation stage, rSLPA is more than two times as fast as SLPA. Considering that rSLPA runs for $200$ iterations while SLPA runs for $100$ iterations, SLPA is over five times more than rSLPA in terms of running time per iteration. This result is consistent with our analysis in Section \ref{subsec:post}, that rSLPA propagates less labels than SLPA does in each single iteration. However, in the post-processing stage, SLPA takes much less time than rSLPA does. This is because in SLPA the memberships of each vertex is calculated by simple thresholding, while in rSLPA, complex operations like finding connected components are needed to find communities. Overall, rSLPA is a bit faster than SLPA in terms of the total running time.

\begin{figure}[htbp]
    \centering\vspace{-3ex}
    \includegraphics[width=.6\linewidth, keepaspectratio]{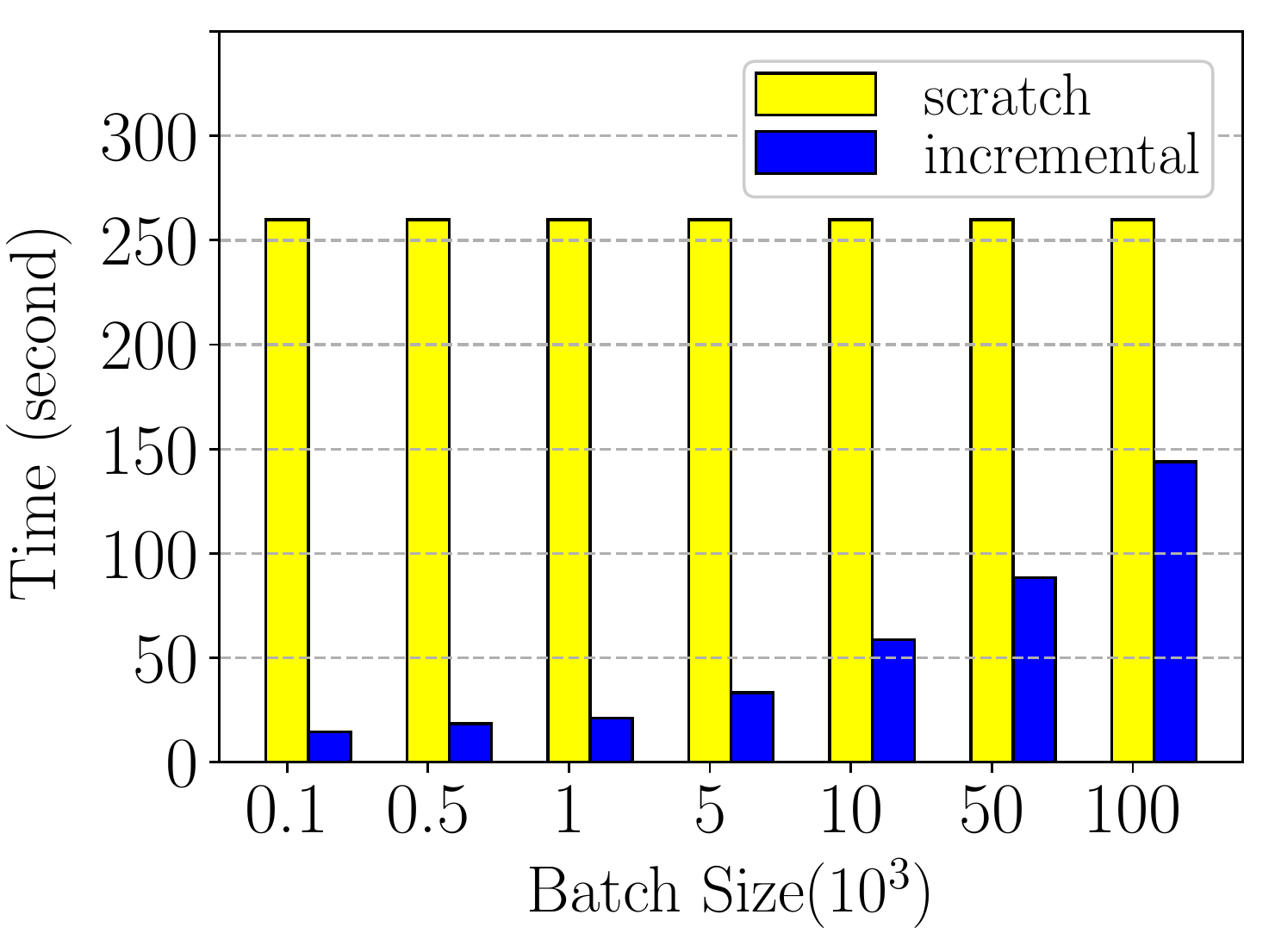}\vspace{-2ex}
    \caption{Running time of rSLPA (incremental updating and calculating from scratch) on different batch sizes.}
    \label{fig:compare_real_dynamic}\vspace{-3ex}
\end{figure}

Figure \ref{fig:compare_real_dynamic} compares the running time of
incremental updating and running from scratch of rSLPA on different
batch sizes. The results show that rSLPA can perform efficient incremental updating. The
increase of running time is sublinear to the increase of batch size.
This is because multiple edges can have the same influence on a
single label, and this overlapping is more frequent with more edges changed.
This also implies that rSLPA is suitable for edit batches of large
size.

\vspace{-1ex}
\subsection{Summary of Experimental Results}
We evaluate the effectiveness and efficiency of rSLPA on synthetic and real-world datasets, respectively. Experiments on synthetic datasets generated by the LFR benchmark show that, rSLPA can detect communities of good quality. The NMI score is over $0.8$ in most of the situations. Experiments on real-world web graphs show that rSLPA is efficient either running from scratch or updating incrementally. When updating incrementally, the running time is sublinear to the changed edge number, which makes rSLPA suitable for large edit batches.

\vspace{-1ex}
\section{Related Work}
\label{sec:related}
The concept of community in networks was first introduced in \cite{girvan2002community} in 2002. It hasn't been long until people realized that in real world networks, communities are overlapped with each other \cite{palla2005uncovering}. Since then, a variaty of of algorithms were proposed in different areas to identify overlapping communities.

\textbf{Clique Based Algorithms.} The Clique Percolation Method
(CPM) \cite{palla2005uncovering} assumes that communities are formed
by adjacent cliques. It detects communities by identifying cliques
and then group adjacent cliques together. CPM naturally finds
overlapping communites since one vertex can belong to several
cliques at the same time, but its running time highly depends on the
clique size $k$.

\textbf{Label Propagation Algorithms.} The Label Propagation
Algorithm (LPA) \cite{raghavan2007near} simulates the communication
between vertices by sending and updating
labels. Then vertices with the same label are assigned to a community. This
algorithm has the complexity linear to the number of edges in the
graph, but can only detect disjoint communities. Later the
Speaker-Listener Label Propagation Algorithm (SLPA)
\cite{xie2012towards} improves LPA by allowing each vertex to hold
multiple labels, thus can detect overlapping communities.

\textbf{Link Partitioning Algorithms.} Unlike previous discussed
algorithms which group vertices together, link partitioning
algorithms group edges/links together. Ahn et al. \cite{ahn2010link}
use Jaccard index as the similarity between edges and then use
Single-linkage clustering to group edges together. Some works like
\cite{evans2009line} and \cite{DBLP:conf/icde/LimRKJL14} map the
original graph into a line graph, then perform community detection
algorithms on the line graph.

\textbf{Local Expansion Algorithms.} Lancichinetti et al.
\cite{lancichinetti2009detecting} give an algorithm that expand
communities from random seeds to maximize a fitness function
locally. Users can control the size and quality of communities by
setting the resolution parameter $\alpha$. Another algorithm iLCD
\cite{cazabet2010detection} gradually adds edges to an empty graph,
and performs community creation, merging and vertex assignment
during this procedure. This algorithm can detect overlapping
communities by assigning one vertex to multiple communities.

\textbf{Fuzzy Detection.} Fuzzy Detection algorithms calculate a
membership vector for each vertex such that some objective functions are minimized or maximized. Zhang
et al. \cite{zhang2007identification} map the graph into Euclidean
space using spectral clustering, then uses Fuzzy C-Means (FCM) to
assign the soft membership of each vertex. In work \cite{ren2009simple},
Expectation-Maximization (EM) is used to compute the soft
membership. Most of the algorithms in this category
require the number of communities to be decided in advance, making
them less flexible in practice.

\textbf{Similarity Based Algorithms.} Zhang et al.
\cite{zhang2009parallel} propose an algorithm based on similarity
estimation. Firstly a similarity matrix is estimated by comparing
the common neighbors of each pair of vertices from the original
graph. Then, a topology graph can be constructed according to the
similarity matrix, which will be used to calculate a new similarity
matrix in the next iteration. By iteratively updating the similarity
matrix and topology graph until convergency, the communities can be
discovered as connected components in final topology graph.

\vspace{-1ex}
\section{Conclusion}
\label{sec:conclusion} In this paper, we study the problem of
overlapping community detection on distributed and dynamic graphs.
After analyzing the limitations of the highly parallelizable
algorithm SLPA, we propose a new algorithm rSLPA by relaxing the
probability distribution in the label propagation stage. With this
technique, our algorithm can incrementally update the result as the
graph dynamically changes. To our best knowledge, rSLPA is the first
algorithm that can incrementally detect overlapping communities over
distributed and dynamic graphs accurately. We conduct extensive
experiments to confirm the good performance of our approaches on
synthetic/real data.


\vspace{-1ex}


\let\xxx=\bibitem\def\bibitem{\par\vspace{-0.2mm}\xxx}

\bibliographystyle{IEEEtran}
\bibliography{community}

\end{document}